\documentclass[manuscript]{aastex631}

\usepackage{soul}
\usepackage{ulem}
\usepackage{xcolor}
\DeclareRobustCommand{\erase}{\bgroup\markoverwith{\textcolor{red}{\rule[.5ex]{2pt}{0.4pt}}}\ULon}

\shorttitle{EOS Dependence in Proto-Uranian Giant Impact Simulations}
\shortauthors{Murashima \& Sasaki}

\begin{document}

\title{Dependence on the Equation of State in SPH Simulations of Proto-Uranian Disk Formation from a Giant Impact}

\author[0009-0008-6885-8122]{Keiya Murashima}
\affiliation{Graduate School of Advanced Integrated Studies in Human Survivability, Kyoto University \\
Yoshida Nakaadachi-Cho 1, Sakyo-ku, Kyoto 606-8306, Japan}
\email{murashima.keiya.3d@kyoto-u.ac.jp}

\author[0000-0003-1242-7290]{Takanori Sasaki}
\affiliation{Department of Astronomy, Kyoto University \\
Kitashirakawa-Oiwake-cho, Sakyo-ku, Kyoto 606-8502, Japan}

\begin{abstract}

The $98^\circ$ obliquity of Uranus is widely attributed to a giant impact that ejected material and formed a debris disk, which subsequently coalesced into its regular satellites.
Previous Smoothed Particle Hydrodynamics (SPH) studies have yielded inconsistent disk compositions, a discrepancy often linked to the variety of numerical and physical modeling assumptions.
We address this by presenting SPH simulations that systematically test three distinct EOS models alongside two SPH schemes (standard SPH, and the enhanced density-independent SPH).
We utilized a $3M_{\oplus}$ impactor and explored a range of impact parameters which are capable of reproducing Uranus’s current spin state.
Our primary finding is that for impacts capable of reproducing Uranus's current rotation, the choice of EOS or SPH scheme barely affects macroscopic features such as the post-impact rotation period, disk mass, or disk size; these properties are primarily controlled by the impact's angular momentum.
In contrast, the disk's rock fraction is highly EOS-dependent.
Our results clarify that while disk mass and size are robust outcomes, the final disk composition is highly model-dependent. 
Therefore, accurate EOS modeling, integrated with detailed disk evolution studies, is essential to definitively validate the giant impact scenario for Uranus.

\end{abstract}

\section{Introduction} \label{sec:intro}

The 98$^\circ$ obliquity of Uranus and the similarly tilted orbits of its major satellites remain a long-standing mystery in planetary science. 
Uranus is estimated to contain approximately 10 wt\% gas \citep{helled2010, podolak1987}, likely accreted from the protoplanetary disk.
One hypothesis for satellite formation suggests that they accreted from ``satellitesimals'' within a circum-Uranian disk \citep[e.g.,][]{szulagyi2018, rufu2022coaccretion+}. 
If, however, satellites formed in a gaseous circumplanetary disk—whose orbital plane would typically align with that of the protoplanetary disk—then both Uranus's spin axis and its satellite orbits must have been tilted by some subsequent mechanism. 
Although spin–orbit resonance can adiabatically tilt the system, achieving a 98$^\circ$ obliquity has proven difficult \citep{boue2010, rogoszinski2020}. 
Moreover, if Uranus's spin axis were abruptly tilted by a collision, the satellite orbits could not adiabatically follow. 
\citet{morbidelli2012} proposed a scenario in which the precession of satellite orbits due to Uranus's tilted $J_2$ component generates a thick torus in the satellite plane; subsequent collisional damping could then realign the orbits with Uranus's equator. 
A major difficulty with this model, however, is that the damped satellite orbits would tend to become retrograde. 
Although they suggested multiple collisions as a possible solution, the detailed mechanism remains unclear. 

\citet{crida2012formation} proposed an alternative scenario in which satellites form from a disk of solid material around Uranus. 
Their analytical model considered a disk extending to the Roche radius, where the planet's tidal force prevents the direct aggregation of solids. 
In such a disk, tidal torques would transport solid components beyond the Roche radius, enabling satellite accretion. 
The newly formed satellites would then migrate outward under tidal torques from both the planet and the disk. 
This iterative process of formation and outward migration would continue until the disk dissipated, producing a full satellite system. 
A key feature of this model is that the disk mass decreases as satellites form, naturally yielding smaller inner satellites—consistent with the present Uranian system. 
However, this scenario requires an independent origin for the solid debris disk itself. 
While a possible source could be the tidal disruption of a protoplanet or comet passing near Uranus, such a mechanism does not guarantee a disk consistent with the observed system \citep[e.g.,][]{hyodo_ring_2017}.

A third scenario involves the formation of satellites from a debris disk generated by a single giant impact that simultaneously tilted Uranus's spin axis. 
In this case, the debris disk would naturally align with Uranus's new equatorial plane, facilitating the formation of orbiting satellites. 
Yet this model also faces major challenges. 
Numerical simulations have shown that oblique impacts of bodies with masses $\sim 1$–$3\,M_{\oplus}$ ($0.07$–$0.2\,M_{\rm U}$, where $M_{\rm U}$ is Uranus's mass) can reproduce Uranus's present spin angular momentum \citep{slattery1992, kegerreis2018, reinhardt2020, woo2022}. 
The impactor–target mass ratios in these simulations are comparable to those inferred for Earth–Moon formation \citep[e.g.,][]{canup2001}, where a lunar-mass body ($\sim 0.01\,M_{\oplus}$) accreted just outside the Roche limit \citep[e.g.,][]{Ida1997} and subsequently migrated outward due to tidal evolution. 
Similarly, proto-Uranus impact simulations typically yield debris disks with masses of a few percent of Uranus's mass, confined mostly within the Roche limit \citep{slattery1992, kegerreis2018, reinhardt2020}. 
This differs starkly from Uranus's current major satellites, whose combined mass is only $\sim 10^{-4}\,M_{\rm U}$. 
Furthermore, whereas Uranus's Roche limit is $\sim 2.1r_{\rm U}$ (with $r_{\rm U}$ Uranus's physical radius), its five largest satellites orbit at distances of 5–25 $r_{\rm U}$. 
Thus, the massive and compact nature of disks predicted by single-impact scenarios poses a serious inconsistency with the present Uranian satellite system \citep{morbidelli2012, ishizawa2019can}.

In response to these challenges, \citet{ida2020} proposed that debris disks generated by impacts on proto-Uranus would be entirely vapor disks, owing to the relatively low evaporation temperature of water ice. 
However, as noted in their study and subsequent works, the rocky component is expected to solidify much more quickly than the icy and volatile components due to its significantly higher condensation temperature \citep{ida2020, woo2022}.
They argued that such disks would undergo substantial viscous evolution until ice grains eventually recondensed. 
By modeling the thermodynamic evolution of the post-impact circum-Uranian disk, they examined the time-dependent surface density distribution of the solid component that could serve as the source material for satellite formation. 
Their results showed that the maximum radius, total mass, and power-law slope of the recondensed ice disk are broadly consistent with the present Uranian satellite system. 
This finding suggests that the system could, in principle, be explained by a giant impact scenario, provided that the thermodynamic evolution of the vapor disk is properly taken into account. 

Nevertheless, several important issues remain unresolved. 
In particular, \citet{ida2020} did not address the composition of the resulting satellites in detail. 
Building on this, \citet{woo2022} conducted a highly comprehensive study that directly coupled SPH impact simulations with disk evolution and N-body accretion models, demonstrating that the masses, compositions, and orbits of the regular satellites can indeed be consistent with an impact origin.
Observations indicate that the five major Uranian satellites contain a rock-to-total-mass (rock + ice) ratio of approximately 30--50\% \citep{jacobson1992masses}. 
By contrast, the rocky material predicted in disks from previous giant impact simulations on Uranus \citep{slattery1992, kegerreis2018, reinhardt2020, woo2022} varies considerably. 
These discrepancies likely arise from differences in the adopted equations of state (EOS) and initial conditions and the resulting initial conditions (e.g., the pre-impact internal structure of the bodies) (Table \ref{table:PreviousStudy}). 
EOS models applied to rock and ice have included ANEOS \citep{thompson1974}, SESAME \citep{holian1984}, the Tillotson EOS \citep{tillotson1962}, and the model by \citet{hubbard1980} (hereafter HM80). 
Because giant impacts on Uranus are expected to generate fully vaporized disks, it is essential to employ EOS formulations that realistically capture phase transitions and accurately track the behavior of low-abundance rocky components within the vapor. 
Furthermore, because the generated circumplanetary disks are generally low in mass and the rock component is in very low abundance, achieving sufficient numerical resolution is equally crucial to accurately resolve the distribution of these materials.

\begin{table}[t]
\caption{Initial conditions of previous studies: H-REOS.3 and He-REOS.3 are ab initio EOS for hydrogen and helium \citep{becker2014}. Details of ``Density correction'' is described in Section 2.1 of \citet{reinhardt2020}}
\begin{center}
\scalebox{0.65}[0.7]{
\begin{tabular}{lcccc}
\hline 
& \citet{slattery1992} & \citet{kegerreis2018} & \citet{reinhardt2020} & \citet{woo2022} \\ \hline \hline
EOS &
\begin{tabular}{c}
ANEOS\\(Iron, Dunite),\\SESAME\\($\mathrm{H_2 O,\,CH_4,\,H_2,\,He}$)
\end{tabular}
&HM80 EOS&
\begin{tabular}{c}
Tillotson EOS\\(Granite, Ice),\\Ideal Gas($\mathrm{\,H_2,\,He}$)
\end{tabular}
&
\begin{tabular}{c}
ANEOS\\(Dunite, Water),\\H-REOS.3($\mathrm{H_2}$),\\He-REOS.3($\mathrm{He}$)
\end{tabular}
\\ \hline
Composiotion&
\begin{tabular}{c}
Inner Core(Iron),\\ Outer Core(Dunite),\\ Envelope($\mathrm{H_2 O,\,CH_4,\,H_2,\,He}$)
\end{tabular}
&
\begin{tabular}{c}
Core(Iron + Dunite),\\Mantle(Ice),\\ Envelope($\mathrm{H_2 O,\,CH_4,\,NH_3}$)
\end{tabular}
&
\begin{tabular}{c}
Core(Granite),\\Mantle(Ice),\\ Envelope($\mathrm{\,H_2,\,He}$)
\end{tabular}
&
\begin{tabular}{c}
Core(Dunite),\\Mantle(Ice),\\ Envelope($\mathrm{\,H_2,\,He}$)
\end{tabular}
\\ \hline
Mass ratio&
\begin{tabular}{c}
Iron/(Iron+Dunite) = 0.31,\\Ice/(Iron+Dunite) = 2.71,\\Gas($\mathrm{H_2,\,He}$) = 2.0 $\mathrm{M_{\oplus}}$
\end{tabular}
&
\begin{tabular}{c}
Core = 2.02 $\mathrm{M_{\oplus}}$,\\Mantle = 11.68 $\mathrm{M_{\oplus}}$,\\Gas($\mathrm{H_2,\,He}$) = 0.84 $\mathrm{M_{\oplus}}$
\end{tabular}
&
\begin{tabular}{c}
Core:Mantle = 1:9,\\Gas($\mathrm{H_2,\,He}$) = 2.0$\mathrm{M_{\oplus}}$
\end{tabular}
&
\begin{tabular}{c}
Core:Mantle = 1:9,\\Gas($\mathrm{H_2,\,He}$) = 2.0$\mathrm{M_{\oplus}}$
\end{tabular}
\\ \hline
Scheme&Standard SPH&Standard SPH&
\begin{tabular}{c}
ISPH+\\ Interface correction
\end{tabular}
& 
\begin{tabular}{c}
ISPH+\\ Interface correction
\end{tabular}\\ \hline
\end{tabular}
}\label{table:PreviousStudy}
\end{center}
\end{table}

In addition to the choice of EOS, the reliability of giant impact simulations also depends on the numerical formulation of SPH. 
The commonly used Standard Smoothed-Particle Hydrodynamics (SSPH) scheme is known to struggle with contact discontinuities and free surfaces, leading to inaccurate treatment of hydrodynamical instabilities \citep[e.g.,][]{okamoto2003, agertz2007, valcke2010, mcnally2012}. 
This limitation arises because SSPH computes the density through a density sum over neighboring particles, which inevitably leads to the artificial over-smoothing of sharp density discontinuities.
At contact discontinuities, where density is inherently non-differentiable, SSPH tends to overestimate density on the low-density side and underestimate it on the high-density side. 
As a result, pressure gradients near these interfaces are miscalculated, producing unphysical repulsive forces. 
Such artifacts are particularly problematic in giant impact simulations, where sharp density jumps occur both at the core–mantle boundary and at planetary surfaces \citep[e.g.,][]{hosono2016}. 
Various alternative SPH formulations and interface corrections have been proposed to mitigate these boundary issues and improve the treatment of multiphase fluids in planetary collisions \citep[e.g.][]{yamamoto2015, wadsley2017, Reinhardt2017, Deng2019a, Deng2019b, reinhardt2020, RuizBonilla2022, pearl2022}.
Among them, the Density Independent SPH (DISPH) method \citep{saitoh2013, hosono2013a, hopkins2013} offers a promising alternative that better handles discontinuities. 
While DISPH significantly improves the treatment of internal contact discontinuities between different materials, it should be noted that it still faces inherent challenges in accurately modeling free surfaces.

In this study, we employ both SSPH and DISPH, together with multiple EOS models (ANEOS/SESAME, Tillotson EOS, and HM80 EOS), to systematically investigate the extent to which these factors contribute to discrepancies in previous simulations. 
Through this approach, we aim to clarify the physical conditions required for the formation of a Uranian satellite system consistent with observations.

\section{Methods} \label{sec:methods}

We employed two Smoothed-Particle Hydrodynamics (SPH) schemes: the Standard SPH (SSPH) and the Density Independent SPH (DISPH). 
Our simulations adopted the Wendland C6 kernel \citep{wendland1995}, the artificial viscosity formulation by \citet{monaghan1992}, the Balsara switch \citep{balsara1995}, and the Barnes--Hut tree algorithm \citep{barnes1986}. 
Efficient parallel computations were carried out using the Framework for Developing Particle Simulator (FDPS) \citep{iwasawa2016}. 

In our SPH formulation, rather than fixing the number of neighbor particles via a variable smoothing length, we adopt a fixed interaction search radius for each particle.
Within this defined range, particles interact via the Wendland C6 kernel.
We selected this specific interaction range and the C6 kernel primarily as a conservative choice to ensure numerical stability across all EOS models and to prevent potential pairing instability, which can otherwise occur when handling extreme density contrasts with the DISPH scheme.
To verify that this relatively broad interaction range does not bias our main findings, we performed preliminary tests using the Wendland C2 kernel with a narrower search radius for the Tillotson EOS.
These tests confirmed that the macroscopic outcomes, such as the total disk mass and the rock-to-ice ratio, are highly consistent regardless of the kernel choice.

\subsection{Equation of State}

Hydrodynamical simulations of giant impacts require an Equation of State (EOS) that can reliably describe planetary materials across wide ranges of temperature, pressure, and density. 
In this study, we employed three distinct EOS models.
The relevant input parameters for the EOS models adopted in our simulations are summarized in Table~\ref{tab:tillotson_params} and ~\ref{tab:other_eos}.
Table \ref{tab:tillotson_params} lists the analytical constants used for the Tillotson EOS, while Table \ref{tab:other_eos} details the specific tabular data and analytical formulations utilized for the ANEOS, SESAME, and HM80 models.

\begin{deluxetable*}{lccccccccccc}
\tablecaption{Parameters for the Tillotson EOS \label{tab:tillotson_params}}
\tablewidth{0pt}
\tablehead{
\colhead{Material} & \colhead{$\rho_0$} & \colhead{$A$} & \colhead{$B$} & \colhead{$E_0$} & \colhead{$E_{iv}$} & \colhead{$E_{cv}$} & \colhead{$a$} & \colhead{$b$} & \colhead{$\alpha$} & \colhead{$\beta$} & \colhead{Reference} \\
\colhead{} & \colhead{[g cm$^{-3}$]} & \colhead{[GPa]} & \colhead{[GPa]} & \colhead{[MJ kg$^{-1}$]} & \colhead{[MJ kg$^{-1}$]} & \colhead{[MJ kg$^{-1}$]} & \colhead{} & \colhead{} & \colhead{} & \colhead{} & \colhead{}
}
\startdata
Granite & 2.68 & 18.0 & 18.0 & 4.87 & 3.50 & 18.0 & 0.5 & 1.3 & 5.0 & 5.0 & \citet{benz1986} \\
Ice (H$_2$O) & 0.917 & 9.47 & 9.47 & 10.0 & 0.773 & 3.04 & 0.3 & 0.1 & 10.0 & 5.0 & \citet{benz1999} \\
\enddata
\tablecomments{$\rho_0$ is the reference density, $A$ and $B$ are bulk moduli constants, $E_0$ is the reference specific internal energy, $E_{iv}$ and $E_{cv}$ are the specific energies of incipient and complete vaporization, respectively, and $a$, $b$, $\alpha$, and $\beta$ are dimensionless constants.}
\end{deluxetable*}

\begin{deluxetable*}{llll}
\tablecaption{Tabular and Analytical EOS Models for ANEOS, SESAME, and HM80 \label{tab:other_eos}}
\tablewidth{0pt}
\tablehead{
\colhead{EOS Model} & \colhead{Material} & \colhead{Table ID / Formulation} & \colhead{Reference}
}
\startdata
ANEOS & Dunite (Mantle) & M-ANEOS (dunite) & \citet{collins2014improvements} \\
SESAME & Water (H$_2$O) & SESAME 7150-301 & \citet{holian1984} \\
HM80 & H, He Mixture & Analytical (Perturbed Polytrope) & \citet{hubbard1980} \\
\enddata
\tablecomments{The tabular EOS models evaluate thermodynamic quantities via bilinear interpolation on pre-computed $(\rho, T)$ grids. The SESAME tables are provided by the Los Alamos National Laboratory.}
\end{deluxetable*}

\subsubsection{ANEOS/SESAME}

Our model utilizes the M-ANEOS semi-analytical equation of state code \citep{thompson1974, melosh_2007, thompson2019maneos} to represent the dunite core.
The specific M-ANEOS parameters for dunite are adopted from \citet{collins2014improvements}.
This is combined with SESAME tabular data for H$_2$O (table 7150-301), while the H$_2$--He envelope is treated as an ideal gas.
For the Dunite impactor (M-ANEOS), the grid consists of 860 $\times$ 744 points. The water EOS table (SESAME) consists of 80 $\times$ 44 points.
This gridding ensures that thermodynamic derivatives remain stable and that the phase boundaries and mixed-phase regions—such as the liquid-vapor coexistence curves inherently computed by the EOS models—are adequately resolved during the SPH interpolation process.
ANEOS generates EOS data for shock physics simulations by expressing the Helmholtz free energy as the sum of three contributions: (1) a cold term, (2) a thermal term, and (3) an electronic term. 
It accounts for three major phase transitions: melting, vaporization, and a single solid–solid transition. 
Similarly, the SESAME library is structured around (1) a cold curve, (2) an ionic component including cold and thermal contributions, and (3) a thermal electronic term, with treatments of shock data and multiple phase transitions. 

ANEOS is an analytic, Helmholtz free-energy equation of state originally developed for shock-physics applications \citep{thompson1974, melosh_2007, thompsonmaneos2019}. 
It constructs the Helmholtz free energy as the sum of a cold ($T=0$) contribution, an ionic/thermal lattice contribution (often Debye-like at low $T$ and approaching ideal behavior at high $T$), and an electronic term that includes thermal excitation/ionization; thermodynamic quantities are obtained by differentiation of $F(\rho,T)$. 
The multiphase formulation allows for solid, liquid, and vapor branches, with mixed-phase regions determined by a Maxwell construction, and optionally a single solid–solid transition calibrated to experimental constraints. 
It should be noted that the current formulation of ANEOS has certain limitations; for instance, it is typically restricted to modeling only two phase transitions simultaneously.
Additionally, its assumption of a constant heat capacity can introduce considerable errors in the thermal component of the free energy, particularly at extreme temperatures \citep{Stewart_2020}thomp.
Material parameters (e.g., bulk modulus, Grüneisen parameter, Debye temperature, high-$T$ electronic terms) are calibrated to static compression and shock Hugoniot data, yielding self-consistent $P(\rho,T)$, $u(\rho,T)$, and phase boundaries across a wide dynamic range.

For water, we use the SESAME library (H$_2$O table 7150; version 7150-301), which provides a standardized, computer-readable set of thermodynamic tables for pressure, energy, entropy, and sound speed over broad $(\rho,T)$ ranges, including sub- and super-critical regions and mixed-phase treatments based on library-wide conventions \citep{lyon_sesame_1992}.
SESAME tables are compiled and vetted at Los Alamos from experimental and theoretical sources and distributed with consistent interpolation routines, facilitating robust hydrocode usage and reproducibility.

Practically, our ANEOS (dunite) and SESAME (water) tables are sampled on $(\rho,T)$ grids and bi-linearly interpolated during SPH updates; the H$_2$--He envelope follows an ideal-gas closure and does not participate in condensed-phase transitions. 

This method is selected for its computational simplicity and numerical robustness, ensuring efficient EOS lookups during the SPH steps.
Given the resolution of the tables utilized in this study, bi-linear interpolation is expected to provide sufficient accuracy while effectively avoiding potential overshoots or unphysical oscillations that can sometimes be associated with higher-order schemes near phase boundaries.
This hybrid approach mirrors prior impact studies of silicate–ice bodies, where ANEOS has been widely used for silicates/iron in Moon-forming impacts \citep[e.g.][]{canup_lunar_2004} and SESAME for water/ice in H$_2$O-rich impacts \citep[e.g.][]{slattery1992, gisler2004two}.
While tabular models such as SESAME and ANEOS account for complex multiphase states, they are constrained by the resolution of their underlying grids, where low-order interpolation can occasionally introduce thermodynamic inconsistencies.

\subsubsection{Tillotson EOS}

The second model employed the Tillotson EOS, supplemented by an ideal gas EOS for the H$_2$–He envelope.  
The rocky component is represented using the granite model of \citet{benz1986}, while the water ice component follows the parameterization of \citet{benz1999}.  

Originally developed in \citet{tillotson1962}, the Tillotson EOS is designed for simplicity and numerical robustness under extreme conditions, such as high pressures and rapid compression or expansion.  
It effectively interpolates between a cold‐curve (low internal energy) regime approximated by a Mie–Grüneisen form and a high‐energy regime treated more similarly to an ideal gas.  
In compressed states below the incipient vaporization energy, it uses a form similar to Mie–Grüneisen, whereas in high energy (post‐shock) or expanded states, the pressure is gradually transferred toward a linear dependence on internal energy (ideal‐gas‐like) to represent vaporization.  

However, the Tillotson EOS has well-known limitations.
Specifically, its simplified modeling of expanded states can lead to a faulty release path during decompression. Furthermore, it lacks thermodynamic completeness; it does not provide entropy, and the temperature must be estimated under simplifying assumptions, which can be crude and inaccurate \citep[e.g.][]{wissing_hobbs_2020}.
It also does not explicitly include latent heat for phase transitions — i.e., there is no thermodynamic treatment of the energy absorbed or released during melting or vaporization.  
Likewise, the mixed-phase region (solid/liquid or liquid/vapor) is not treated via rigorous equilibrium constructions (e.g., Maxwell constructions) or via separate free‐energy branches.  
Consequently, thermodynamic quantities such as entropy and detailed vapor fraction may be inaccurate in regions near phase boundaries.  

Despite these limitations, the Tillotson EOS remains widely used in studies of planetary collisions and giant impacts \citep[e.g.][]{benz_moon_impact_1989, reinhardt2020} because of its computational efficiency, its relative simplicity (fewer material parameters to calibrate), and its ability to produce qualitatively plausible behavior over wide ranges of density and internal energy \citep[e.g.][]{brundage_2013, wissing_hobbs_2020, meier_etal_2021}.
In our study, we exploit these advantages to compare how this simpler EOS performs relative to more realistic model (ANEOS/SESAME), particularly in predicting rock ejection and disk composition, but keep in mind that quantitative differences near phase transitions are expected.

\subsubsection{HM80 EOS}

The third model adopted the HM80 EOS \citep{hubbard1980} for all major components of proto-Uranus: An iron+rock core, an icy mantle, and a gaseous H$_2$–He envelope. 
The assumed composition consists of a rocky core (SiO$_2$, MgO, FeS, FeO), an icy mantle (H$_2$O, NH$_3$, CH$_4$), and an outer He–H$_2$ envelope. 
The HM80 EOS was originally constructed by estimating Uranus's internal density profile from observational constraints (e.g., mass, radius, and gravitational moments) and then fitting analytic pressure–temperature relations to reproduce those interior models. 
It therefore represents a semi-empirical EOS tailored for ice giant interiors, rather than a general-purpose shock-physics EOS.  

A key feature of HM80 is that pressure is expressed as a function of density and temperature, $P(\rho,T)$, whereas our SPH formulation uses internal energy $u$ as the fundamental thermodynamic variable. 
Following \citet{kegerreis2018}, we therefore convert $u$ to $T$ by
\begin{eqnarray}
u_0(\rho) &=& \int^{\rho}{\rho_0}\frac{P_0(\rho)}{\rho^2}d\rho , \\
u(\rho, T) &=& u_0(\rho) + C_{\rm V} T,
\end{eqnarray}
where $u_0$ and $P_0$ are the specific internal energy and pressure at zero temperature, $\rho_0$ is the zero-pressure density, and $C_{\rm V}$ is the specific heat capacity.  

The main advantage of HM80 is its computational simplicity and its design consistency with Uranus's interior structure models. 
However, it should be emphasized that HM80 does not explicitly incorporate phase transitions such as melting or vaporization. 
As a result, it lacks the ability to capture the thermodynamics of mixed-phase states and may under- or over-estimate entropy in strongly shocked or vaporized regimes. 
This limitation is particularly relevant for giant impact simulations, where disks are expected to be largely or entirely vaporized. 
In addition, it should be noted that extrapolating its fitted formula to the extreme high-temperature and low-density regimes characteristic of giant impacts involves inherent uncertainty.
Accordingly, HM80 serves here as a useful comparative model, but its predictions must be interpreted with caution when assessing rock/ice partitioning in debris disks.

\subsection{Initial Conditions}

\begin{table}[t]
\caption{Model parameters of the simulations. The range of $L_{\rm imp}$ is given in units of $10^{36}\,{\rm kg\,m^2\,s^{-1}}$.}
\begin{center}
\begin{tabular}{cccc}
\hline
Run & EOS & SPH scheme & $L_{\rm imp}$ \\
\hline
SESAME/ANEOS-DI & ANEOS (Dunite) \& SESAME (H$_2$O) \& Ideal gas & DISPH & 1--9 \\
SESAME/ANEOS-S  & ANEOS (Dunite) \& SESAME (H$_2$O) \& Ideal gas & SSPH  & 1--9 \\ 
Tillotson-DI    & Tillotson EOS (Granite, Ice) \& Ideal gas        & DISPH & 1--9 \\
Tillotson-S     & Tillotson EOS (Granite, Ice) \& Ideal gas        & SSPH  & 1--9 \\
HM80-DI         & HM80 EOS (Dunite, Ice, H--He)                   & DISPH & 1--9 \\ 
HM80-S          & HM80 EOS (Dunite, Ice, H--He)                   & SSPH  & 1--9 \\
\hline
\end{tabular}
\label{table:InitialCondition}
\end{center}
\end{table}

We constructed an ice-rich proto-Uranus together with a differentiated impactor.
The impactor is assumed to have the same interior composition as the target body—consisting of an iron-rock core and an H$_2$O ice mantle—but is devoid of a gaseous envelope.
Following \citet{hubbard1980}, the global ice-to-(rock+ice) mass ratio of the system was set to 2.71. 
The proto-Uranus was assumed to possess a $2\,M_{\oplus}$ H–He envelope, consistent with interior models of ice giants. 
The combined mass of the proto-Uranus and the impactor was fixed at $14.5\,M_{\oplus}$, with the impactor mass set to $3\,M_{\oplus}$. 
This corresponds to an impactor-to-target mass ratio of $\sim$ 0.25, within the range suggested by previous Uranus impact scenarios. 
We strictly enforced this identical bulk composition (i.e., the same global mass fractions of core, mantle, and envelope) across all EOS models.
This setup is essential to isolate the dynamic and thermodynamic effects of the chosen EOS from variations in the initial bulk composition.
We note that the internal structure of Uranus remains uncertain due to observational ambiguities in its gravity field, shape, and rotation period \citep[e.g.,][]{helled_uranus_2010, nettelmann2013, morf2024}. 

Each body was discretized into $\sim 10^{5}$ SPH particles, and all SPH particles are assigned an identical mass.
To confirm that our adopted resolution ($\sim 10^5$ particles) do not artificially affect our macroscopic results, we conducted convergence tests utilizing a higher resolution ($\sim 10^6$ particles).
The details and results of these tests are presented in Appendix B, demonstrating the numerical robustness of our calculations.
Particles belonging to the core, mantle, and envelope were each assigned the EOS listed in Table~\ref{table:InitialCondition}. 

We carried out a total of 54 simulations, systematically varying the EOS, the SPH scheme, and the initial angular momentum $L_{\rm imp}$. 
The value of $L_{\rm imp}$ ranged from $1 \times 10^{36}$ to $9 \times 10^{36}\,{\rm kg\,m^2\,s^{-1}}$, corresponding to $0.77$–$7.7$ times Uranus's present spin angular momentum. 
The relative velocity at infinity $v_\infty$ was fixed at 5.0 km\,s$^{-1}$, consistent with previous SPH studies of giant impacts \citep[e.g.][]{kegerreis2018, reinhardt2020}.
Since Uranus's escape velocity is approximately four times its Keplerian orbital velocity, the resulting impact velocity is effectively insensitive to the choice of $v_\infty$.  

In this setup, the system's angular momentum $L_{\rm imp}$ effectively determines the impact parameter: lower $L_{\rm imp}$ corresponds to more head-on collisions, whereas higher $L_{\rm imp}$ corresponds to increasingly oblique encounters. 
Thus, by varying $L_{\rm imp}$, we systematically explore the transition from nearly head-on to grazing impacts without prescribing a fixed impact angle. 

Typical wall-clock runtimes for individual simulations were 3–6 days, with higher $L_{\rm imp}$ requiring longer runtimes.
This increased computational cost for high-angular-momentum (grazing) collisions is due to the extended physical simulation time required; such grazing impacts frequently result in secondary collisions, taking much longer for the widely dispersed material to dynamically settle into a stable circum-planetary disk.

\subsection{Classification of SPH Particles}

To investigate the properties of the debris disks, SPH particles in this study were classified into three categories: ``planetary,'' ``disk,'' or ``unbound,'' following the approach of \citet{reinhardt2020}.
Particles with positive total energy (kinetic + potential) were first identified as ``unbound''.
All remaining bound particles were then divided between the planet and the disk using the iterative algorithm developed by \citet{canup2001} \footnote{An alternative classification method used for comparison is described in Appendix \ref{appendix_A}.}.
For clarity, the main steps of this procedure are summarized below:

\begin{enumerate}
\item Identify the central planet based on the mean density of Uranus and an initial estimate of its radius.
\item Determine all particles that are gravitationally bound to this central object.
\item For each bound particle, compute the specific angular momentum relative to the planet.
Particles whose pericenter distance lies within the planet's physical radius, or whose angular momentum is insufficient to sustain an orbit, are classified as ``planetary'' (i.e., accreted onto the planet).
The remaining bound particles are assigned to the ``disk.''
\item Update the planet's mass and gravitational potential to include the newly accreted material.
\item Repeat steps 1–4 iteratively until the planet–disk mass partition converges.
\end{enumerate}

This self-consistent procedure accounts for the evolving mass and potential of the central planet and minimizes the misclassification of marginally bound particles that could otherwise be ambiguously assigned. For simulations with an impact angular momentum of $9.0 \times 10^{36}\,{\rm kg\, m^2\, s^{-1}}$, the collision is not fully resolved by the end of the calculation, because the impactor survives the secondary impacts within 144 hours. Therefore, particles gravitationally bound to the surviving impactor are excluded from the analysis.

\section{Results} \label{sec:results}

\subsection{Typical Results}

\begin{figure}[ht!]
\includegraphics[width=\linewidth]{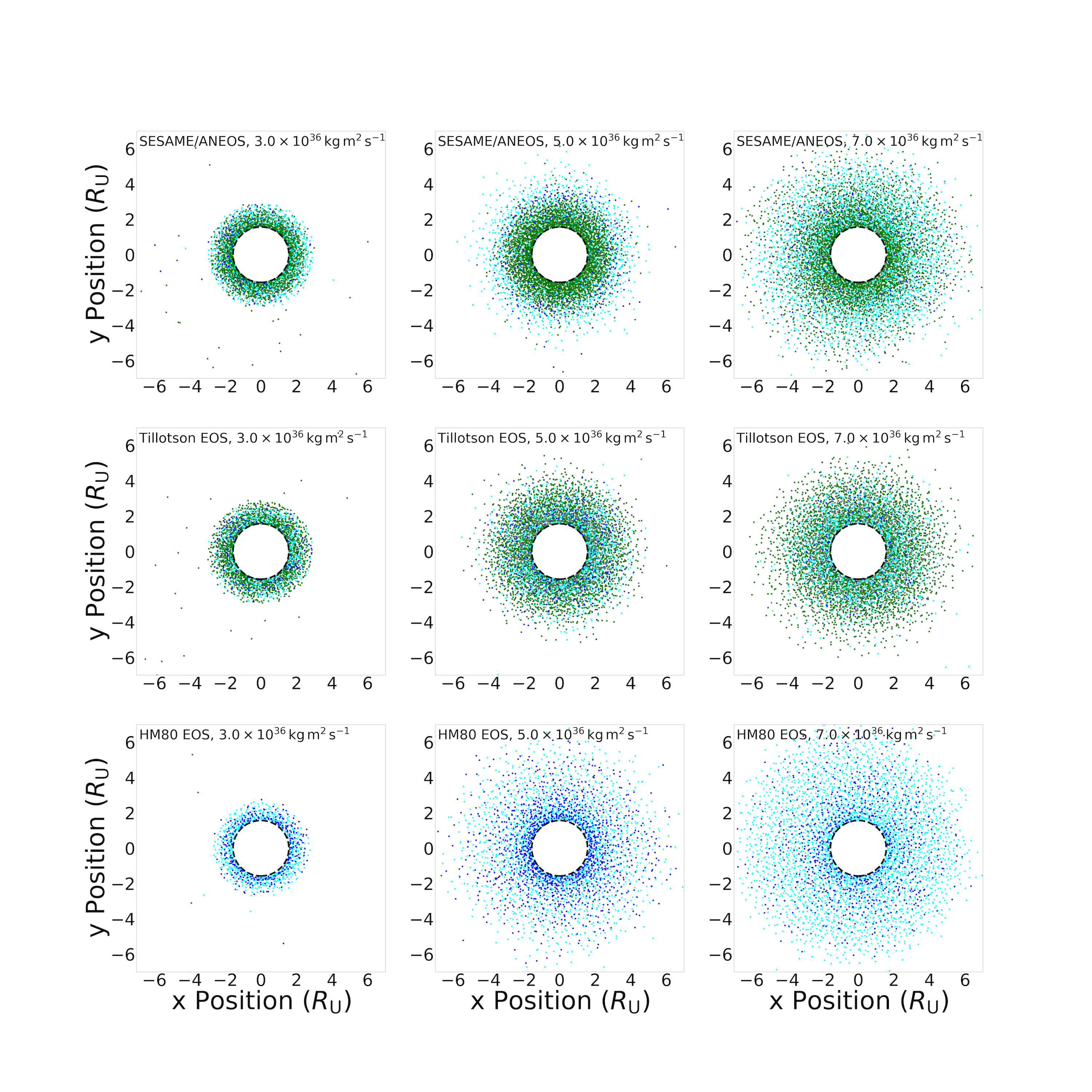}
\caption{Particles ejected by collisions simulated with the DISPH method using SESAME/ANEOS EOS (top), Tillotson EOS (middle), and HM80 EOS (bottom), for initial angular momenta $L_{\rm imp} = 3.0$, $5.0$, and $7.0 \times 10^{36}\, {\rm kg\,m^2\,s^{-1}}$ (from left to right). These panels display the state of the system at the final time of the simulation. The dotted circle indicates Uranus's Roche radius. The components are color-coded as follows: Target ice (dark blue), impactor ice (light blue), impactor rock (gray), and atmosphere (green).}\label{figure:snapshots}
\end{figure}

\begin{figure}[ht!]
\includegraphics[width=\linewidth]{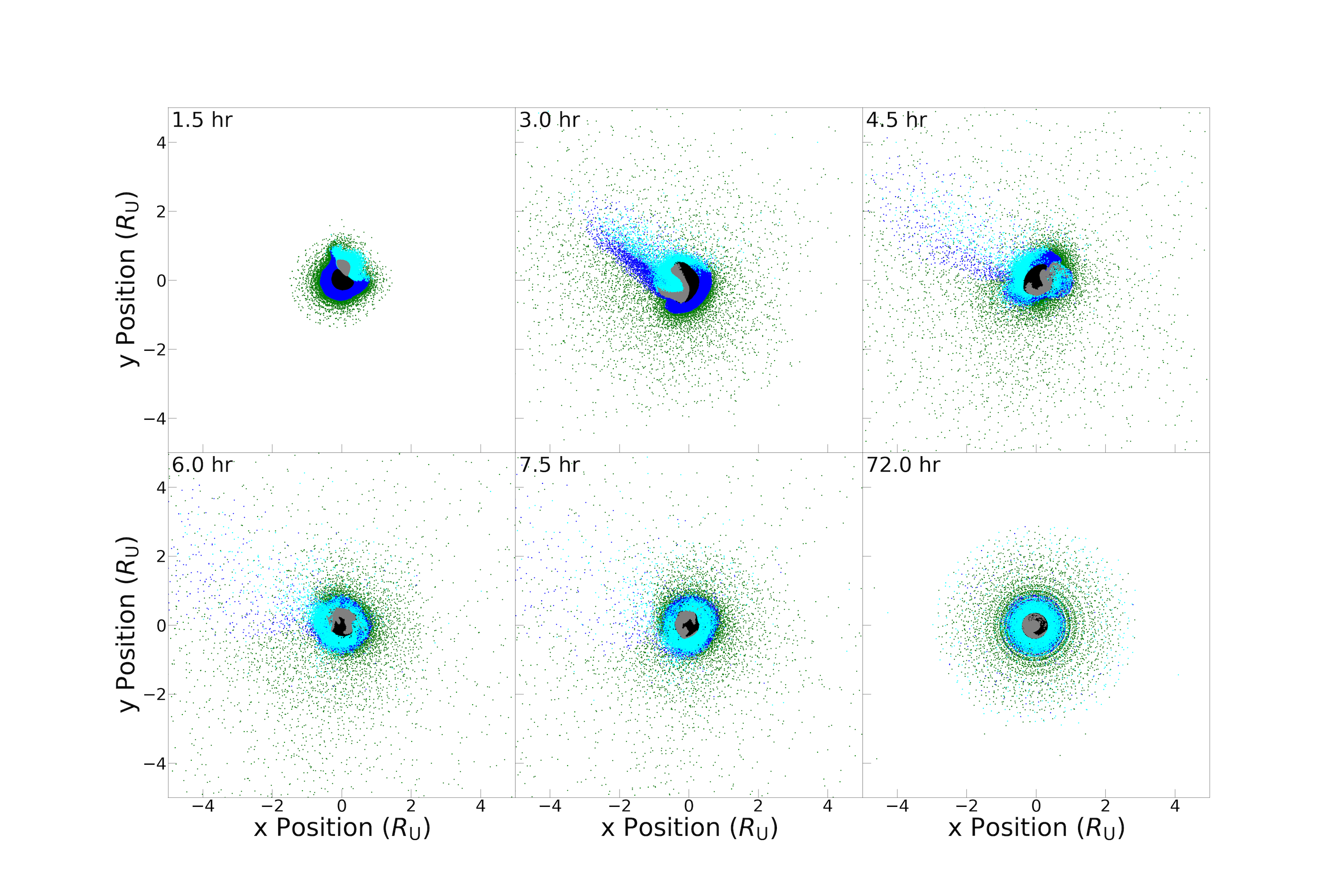}
\caption{Snapshots of a collision with $L_{\rm imp} = 3 \times 10^{36}\, {\rm kg\,m^2\,s^{-1}}$ using the Tillotson EOS. 
Only particles with $z < 0$ are shown. Snapshot times are measured in hours from the beginning of the simulation.}\label{figure:snapshot_SESAME_DI_3}
\end{figure}

\begin{figure}[ht!]
\includegraphics[width=\linewidth]{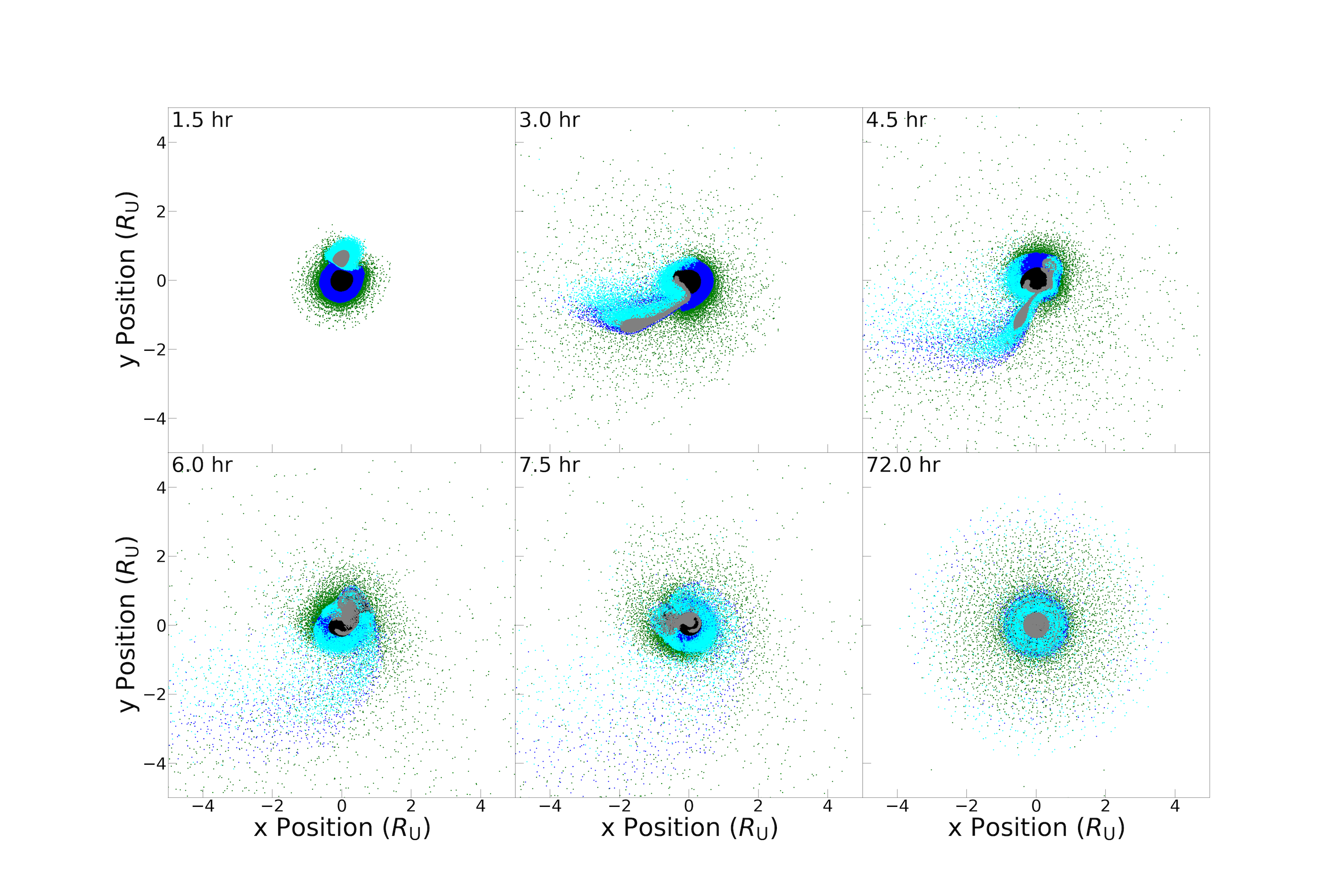}
\caption{Same as Fig.~\ref{figure:snapshot_SESAME_DI_3}, but for $L_{\rm imp} = 5 \times 10^{36}\, {\rm kg\,m^2\,s^{-1}}$.}\label{figure:snapshot_SESAME_DI_5}
\end{figure}

\begin{figure}[ht!]
\includegraphics[width=\linewidth]{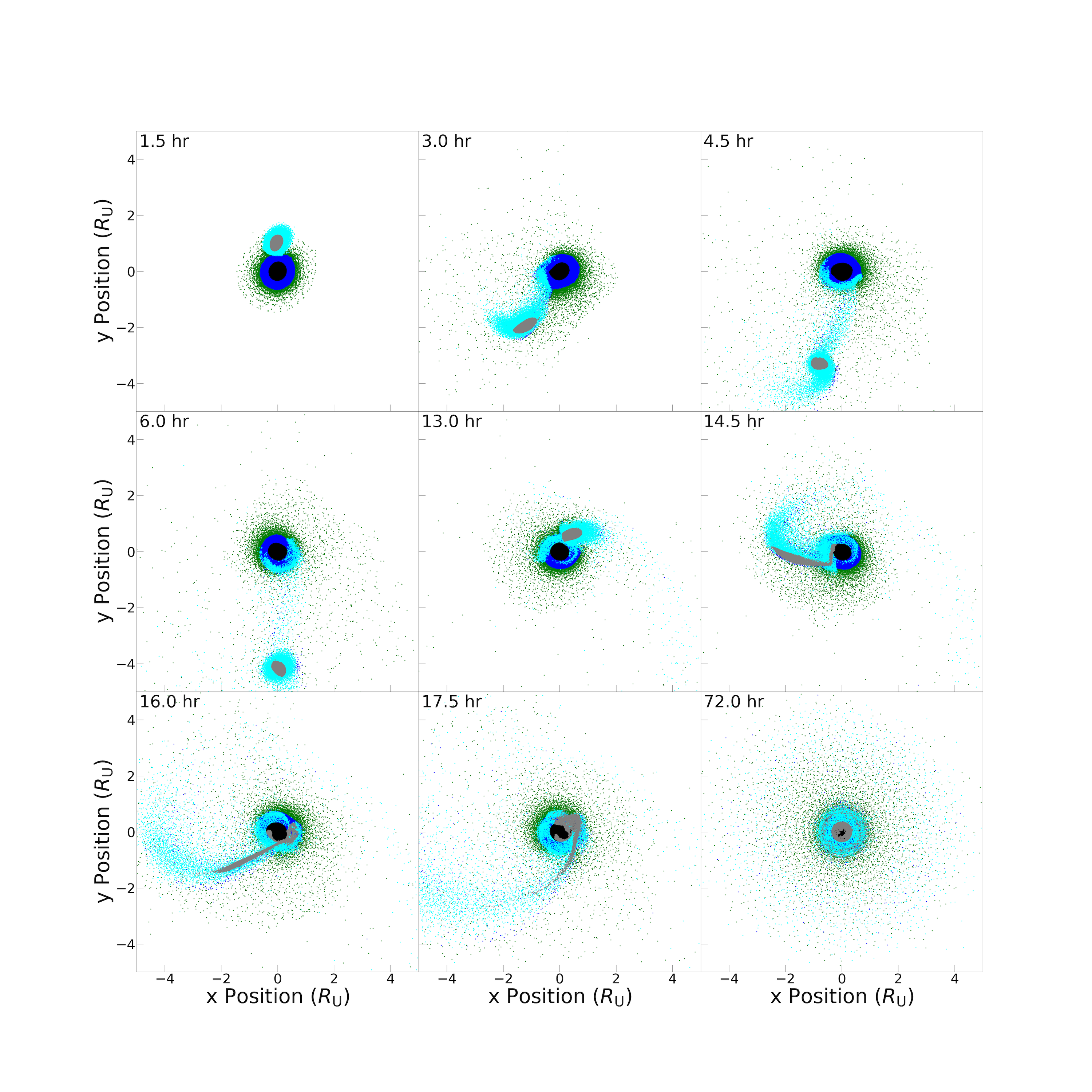}
\caption{Same as Fig.~\ref{figure:snapshot_SESAME_DI_3}, but for $L_{\rm imp} = 7 \times 10^{36}\, {\rm kg\,m^2\,s^{-1}}$.}\label{figure:snapshot_SESAME_DI_7}
\end{figure}

Figure~\ref{figure:snapshots} presents the distribution of particles ejected in collisions simulated with the DISPH scheme under three different EOS models. 
Because the HM80 EOS is primarily calibrated for high-density, high-pressure interior states, it lacks an accurate thermodynamic representation of low-density, highly compressible gases.
Consequently, the H-He envelope particles in the HM80 models behave with artificial cohesion and do not expand or scatter realistically during the impact.
The atmosphere of HM80 EOS is derived by fitting to a highly dense state, so ``atmosphere'' particles have not been scattered by collisions.
Figures~\ref{figure:snapshot_SESAME_DI_3}--\ref{figure:snapshot_SESAME_DI_7} show the temporal evolution of collisions with $L_{\rm imp} = 3.0$, $5.0$, and $7.0 \times 10^{36}\, {\rm kg\,m^2\,s^{-1}}$ using the SESAME/ANEOS EOS by DISPH. 
In general, these impacts produce an arm-like structure through which both the impactor material and the H--He gas of the target are expelled. 
Such spiral-arm features are common outcomes in giant impact simulations of both terrestrial planets \citep[e.g.,][]{canup2001} and ice giants \citep[e.g.,][]{slattery1992, kegerreis2018, reinhardt2020}. 

For collisions with low initial angular momentum ($L_{\rm imp} \lesssim 3 \times 10^{36}\, {\rm kg\,m^2\,s^{-1}}$) (Fig.~\ref{figure:snapshot_SESAME_DI_3}), the impactor is largely assimilated into the target, suppressing the formation of an extensive disk. 
In such cases, the target's rock core merges with the impactor's core, preventing the ejection of rocky material. 
Most of the impactor's ice is likewise incorporated into the target's interior. 
As a result, the debris disk is composed primarily of target ice and H--He gas. 
The total disk mass in these low-$L_{\rm imp}$ cases is typically $\lesssim 10^{-3} M_{\rm U}$, well below the threshold required to form Uranus's present satellites. 

By contrast, in high-angular-momentum collisions ($L_{\rm imp} \gtrsim 7 \times 10^{36}\, {\rm kg\,m^2\,s^{-1}}$) (Fig.~\ref{figure:snapshot_SESAME_DI_7}), the target and impactor cores do not merge upon first contact. 
Instead, an extended arm forms that includes both the rock core and ice mantle of the impactor, thereby injecting a significant rock component into the debris disk. 
Consequently, the disk contains a mixture of target ice and H--He gas together with impactor-derived ice and rock. 
The remainder of the impactor is eventually accreted into the target's interior. 
Because of the higher angular momentum, the total disk mass in these cases is greater than that produced in lower-angular-momentum collisions, reaching $\sim 10^{-2} M_{\rm U}$ in some runs. 
This trend, consistent with previous studies \citep[e.g.,][]{slattery1992, woo2022}, highlights the crucial role of impact geometry in determining whether rocky material is retained in the disk or incorporated into the target.

\subsection{Spin Rotation Periods}

\begin{figure}
\includegraphics[width=\linewidth]{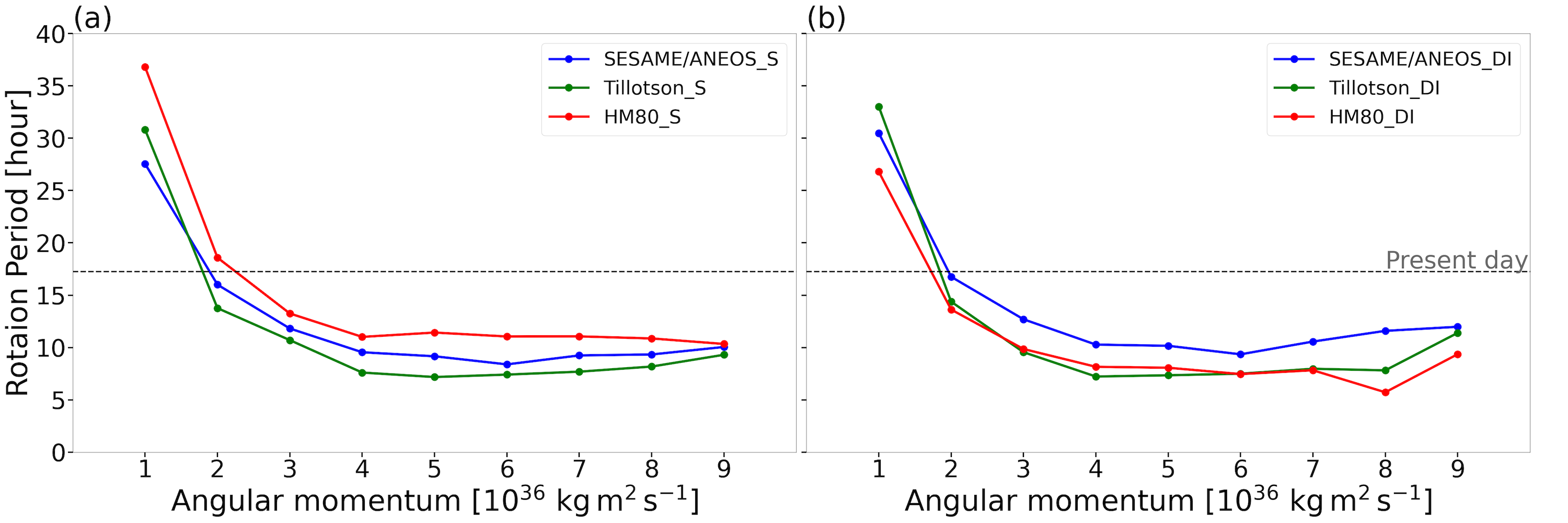}
\caption{Median rotation periods of the post-impact planets as a function of the impact angular momentum $L_{\rm imp}$, for different EOSs and SPH schemes. Panels show (a) SSPH and (b) DISPH. The dashed line indicates the present rotation period of Uranus ($17.24$ hr; \citealt{warwick1986}).}\label{figure:RotationPeriod}
\end{figure}

Figure~\ref{figure:RotationPeriod} shows the rotation periods of the post-impact planets as a function of the initial angular momentum $L_{\rm imp}$. 
The present spin angular momentum of Uranus is estimated to be $L_{\rm U} \simeq (1.28$--$1.33)\times 10^{36}\,{\rm kg\,m^2\,s^{-1}}$ \citep{kaula1968, podolak1987}, corresponding to a rotation period of $17.24$ hr \citep{warwick1986}. 
In our simulations, $L_{\rm imp}$ spans $0.77$--$7.7\,L_{\rm U}$, thereby covering nearly an order of magnitude in angular momentum space. 
It should be noted that some of these grazing impact models deposit a massive excess of angular momentum into the target, resulting in a post-impact planet that rotates significantly faster than the present-day Uranus.
However, this excess angular momentum can potentially be shed during the subsequent long-term evolution, such as through disk-planet interactions, atmospheric mass loss, or the outward migration of the newly formed satellites.

A necessary condition for reproducing the Uranian system is that the simulated post-impact spin period be shorter than 17.24 hr, since excess angular momentum can be transferred from the planet's spin to the orbits of satellites during the subsequent tidal evolution. 
Our results show that this requirement is satisfied in collisions with $L_{\rm imp} \geq 3 \times 10^{36}\,{\rm kg\,m^2\,s^{-1}}$. 
In these cases, the predicted spin periods typically fall in the range $8$–$15$ hr, broadly consistent across the three EOSs and both SPH schemes. 
For lower angular momenta ($L_{\rm imp} \lesssim 2 \times 10^{36}\,{\rm kg\,m^2\,s^{-1}}$), the post-impact planets rotate more slowly ($\gtrsim 20$ hr), which would make it difficult to reproduce the present Uranian spin state without invoking additional angular momentum exchange. 
Note that these rotation periods correspond to the hot, inflated state of the planet immediately after the impact.
Subsequent cooling and contraction will decrease the planet's moment of inertia and thus shorten the rotation period, as the bound angular momentum is conserved.
We present the rotation period here primarily to facilitate direct comparisons with previous giant impact studies \citep[e.g.][]{kegerreis2018, reinhardt2020} and to demonstrate that this macroscopic outcome is largely insensitive to the choice of EOS.

The weak dependence on EOS or SPH scheme indicates that the gross spin state of the planet is primarily controlled by the initial angular momentum of the collision rather than by the thermodynamic treatment of materials. 
This finding is consistent with earlier impact studies of Uranus \citep[e.g.,][]{slattery1992, kegerreis2018, reinhardt2020}, as well as with canonical Moon-forming impacts \citep[e.g.,][]{canup2001}, where the final planetary spin is likewise set mainly by the impact geometry. 
Thus, while EOS choice is crucial for determining the composition of the resulting debris disk, the overall spin evolution of Uranus after a giant impact is relatively robust with respect to the material model.

\subsection{The Mass and Size of the Debris Disk}

\begin{figure}[t]
\includegraphics[width=\linewidth]{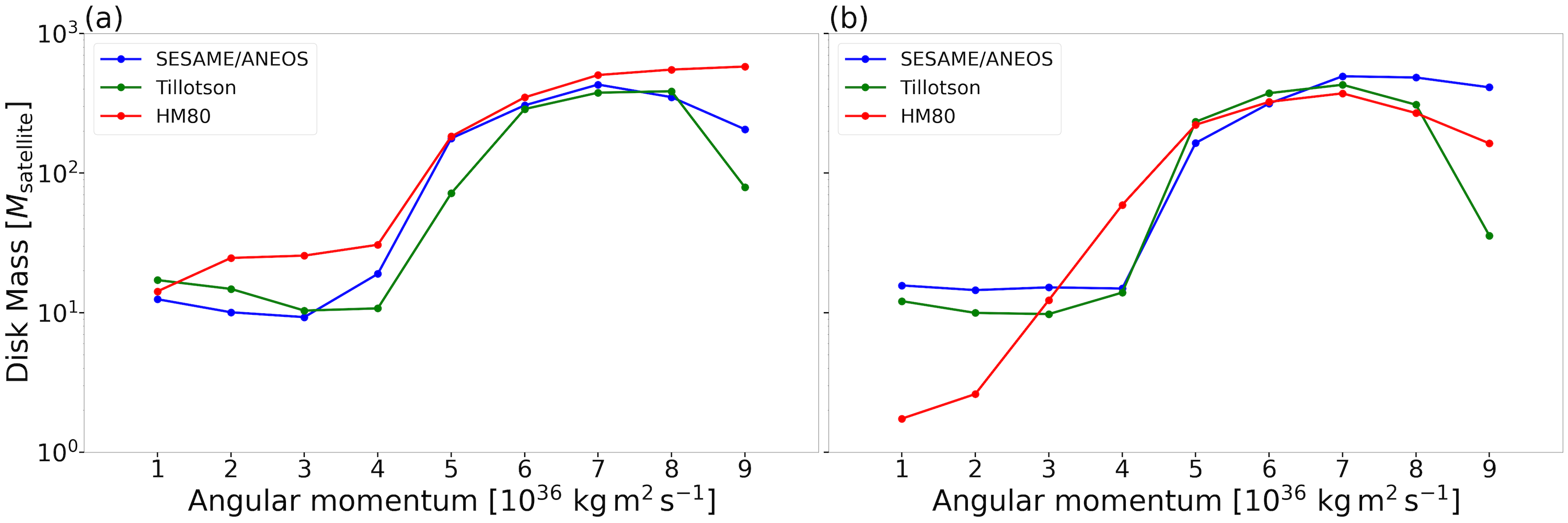}
\caption{Disk mass $M_{\rm d,imp}$ as a function of the initial angular momentum, for (a) SSPH and (b) DISPH. Masses are normalized by the total mass of Uranus's current satellites ($M_{\rm satellite} \sim 10^{-4} M_{\rm U}$). The $y$-axis is shown on a logarithmic scale. }\label{fig:DiskMass}
\end{figure}

\begin{figure}[t]
\includegraphics[width=\linewidth]{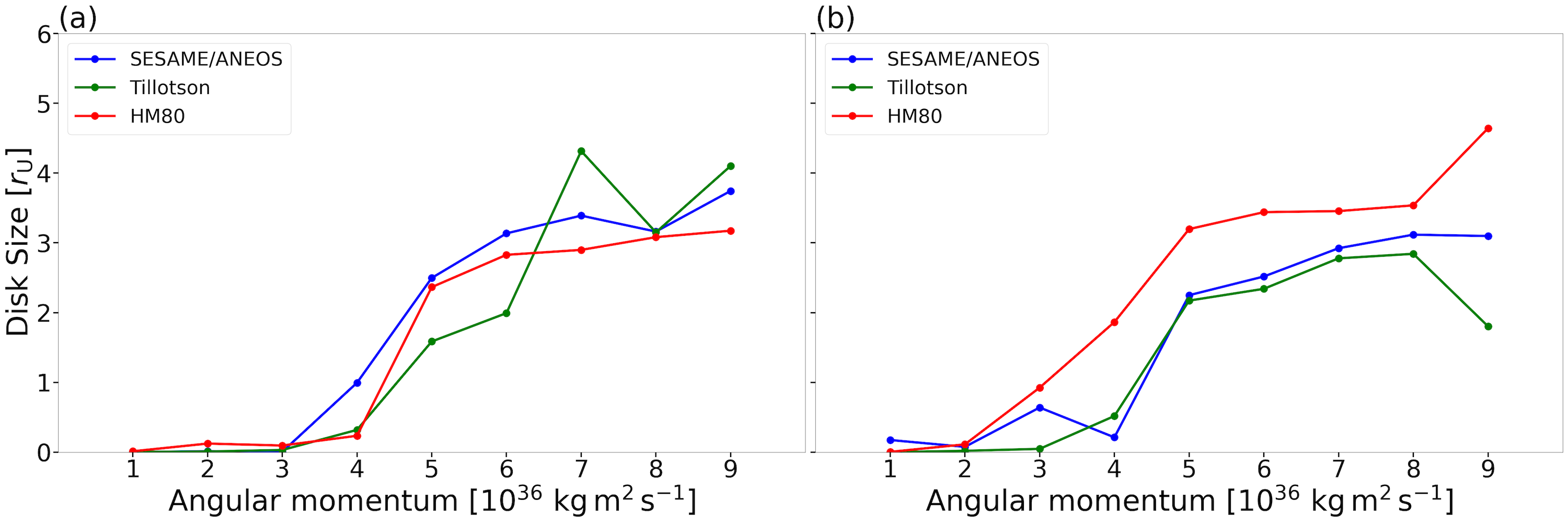}
\caption{Characteristic disk size $\langle r_{\rm d,imp}\rangle$ as a function of the initial angular momentum, for (a) SSPH and (b) DISPH. Sizes are normalized by Uranus's radius. }\label{fig:DiskSize}
\end{figure}

We investigated both the disk mass $M_{\rm d,imp}$ and the characteristic disk size $\langle r_{\rm d,imp}\rangle$, defined as
\begin{equation}
    \langle r_{\rm d,imp}\rangle = 
    \left(\frac{J_{\rm d,imp}}{M_{\rm d,imp}\,r_{\rm U}^2\,\Omega_{\rm U}}\right)^2 r_{\rm U},
\end{equation}
where $J_{\rm d,imp}$ is the total angular momentum of the debris disk and $\Omega_{\rm U}$ is the orbital frequency at $r = r_{\rm U}$. 
This analytical expression, originally proposed by \citet{ida2020}, was also employed as the defining equation by \citet{woo2022}.

Figures~\ref{fig:DiskMass} and \ref{fig:DiskSize} summarize the dependence of $M_{\rm d,imp}$ and $\langle r_{\rm d,imp}\rangle$ on $L_{\rm imp}$. 
Across all EOSs and both SPH schemes, the resulting disks are systematically more massive than Uranus's present satellite system. 
Typical values are $M_{\rm d,imp} \sim 10^{-3}$--$10^{-2} M_{\rm U}$, corresponding to $10^1$--$10^2$ times the combined mass of the existing major satellites. 
This overproduction of disk material is consistent with previous SPH studies of Uranus impacts \citep[e.g.,][]{slattery1992, kegerreis2018, reinhardt2020}, and highlights the need for substantial mass loss during the subsequent viscous and thermodynamic evolution of the disk \citep{ida2020}. 

The characteristic disk sizes are typically confined within a few Uranian radii, with $\langle r_{\rm d,imp}\rangle \sim 2$--$4\,r_{\rm U}$ in most cases. 
This compactness is also consistent with earlier works \citep{slattery1992, kegerreis2018, reinhardt2020}, and implies that outward transport mechanisms are required to explain the current distribution of the major satellites at 5--25 $r_{\rm U}$. 
Such transport could plausibly occur through viscous spreading of a vapor disk \citep{ida2020}, tidal torques between the disk and forming satellites, or subsequent gravitational interactions \citep[e.g.][]{crida2012formation, woo2022}.

Overall, our results indicate that disk mass and size are governed primarily by the impact angular momentum, with only minor variations between EOSs or SPH schemes. 
Thus, while the gross properties of the disk appear robust, the composition of the disk---and in particular the rock-to-ice ratio---is much more sensitive to the adopted physical model, as discussed in the following section.

\subsection{Rock Abundance in the Disk}

\begin{figure}[t]
\includegraphics[width=\linewidth]{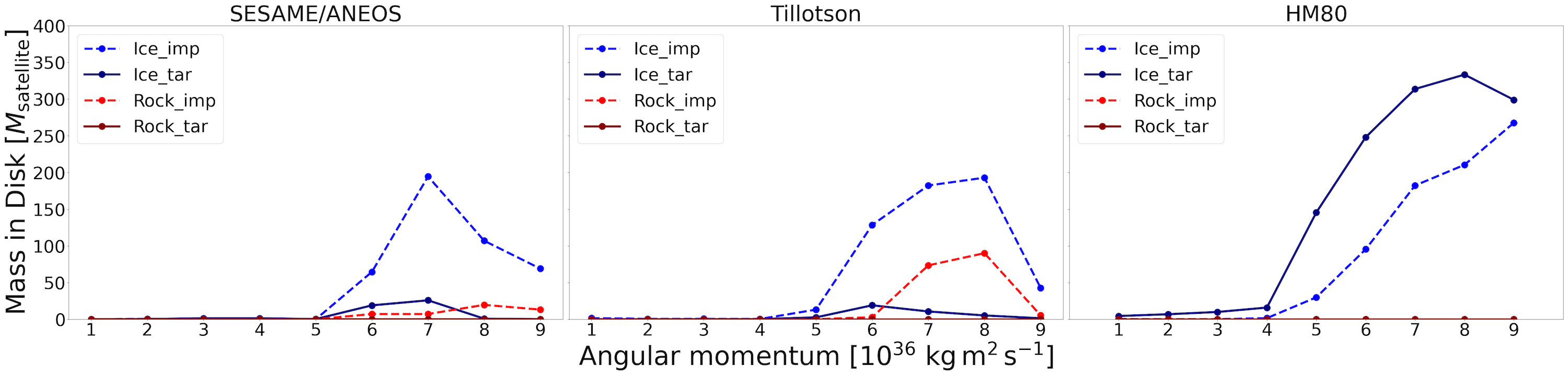}
\caption{Rock and ice components in the debris disk as a function of $L_{\rm imp}$ for different EOSs in SSPH. The total mass of the rock component is normalized by the present satellite mass $M_{\rm satellite}$.}\label{fig:DC_S}
\end{figure}

\begin{figure}[t]
\includegraphics[width=\linewidth]{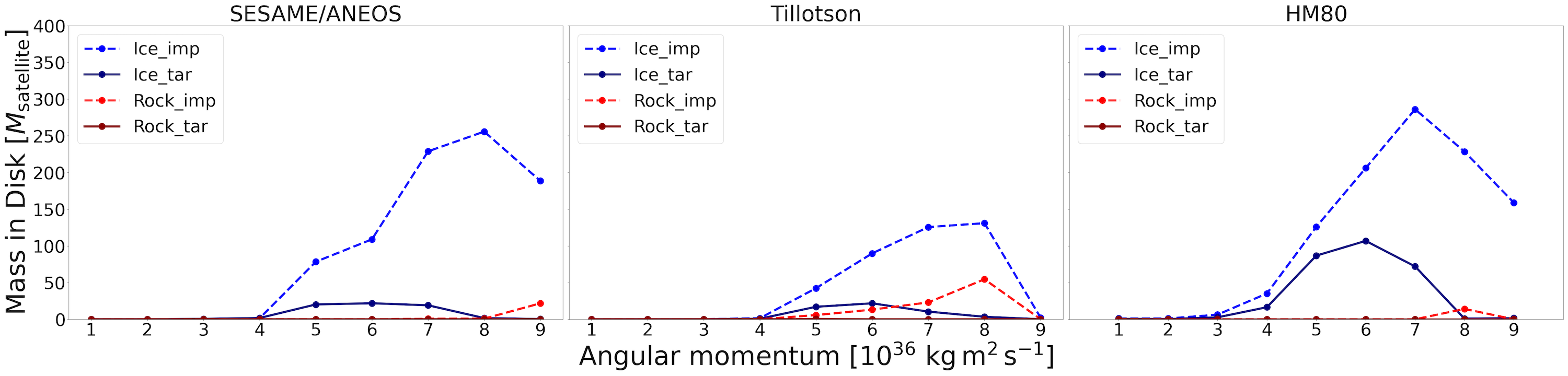}
\caption{Rock and ice components in the debris disk as a function of $L_{\rm imp}$ for different EOSs in DISPH. The total mass of the rock component is normalized by the present satellite mass $M_{\rm satellite}$.}\label{fig:DC_DI}
\end{figure}

Figures~\ref{fig:DC_S} and \ref{fig:DC_DI} show the rock and ice components of the debris disk for different EOSs in DISPH and SSPH, respectively. 
For both the Tillotson and SESAME/ANEOS EOSs, the total rock volume within the disk shows no significant variation between the two SPH schemes, though DISPH tends to yield a smaller rock volume.
In contrast, the ice mass shows a scheme dependence: Switching from SSPH to DISPH increases the ice mass for SESAME/ANEOS but decreases it for the Tillotson EOS. 
For the HM80 EOS, high-angular-momentum collisions exhibit the opposite trend, with DISPH yielding less ice but more rock compared to SSPH. 

Turning to the role of the EOS itself, substantial differences emerge. 
The Tillotson model systematically produces rock-rich disks in high-angular-momentum impacts, with ejection of impactor-core material clearly visible in the simulations. 
In these cases, the rock fraction of the disk can reach $\sim$ 30--40\% of the total disk mass, broadly consistent with the 30--50\% rock content inferred for Uranus's present major satellites \citep{jacobson1992masses, woo2022}. 
By contrast, disks produced with the SESAME/ANEOS EOS typically contain less than a few percent rock, even when the total disk mass is comparable. 
The HM80 EOS yields intermediate results, with a modest increase in rock fraction under oblique, high-angular-momentum impacts, but still generally falling short of the observed satellite composition. 

These differences arise from the EOS-dependent treatment of planetary structure and phase transitions. 
In particular, the Tillotson EOS neglects latent heat, resulting in more efficient disruption and ejection of rocky core material during grazing impacts. 
SESAME/ANEOS, by contrast, includes phase transitions and maintains a stronger density contrast between silicate and ice layers, which favors retention of rocky material within the merged planet. 
HM80, calibrated to Uranus's interior structure, lacks explicit phase changes and therefore lies between the two extremes. 

Overall, our results indicate that the inferred rock fraction of the debris disk is highly sensitive to the adopted EOS, even when the overall disk mass remains comparable. 
This finding echoes the discrepancies among previous Uranus impact studies \citep[e.g.,][]{slattery1992, kegerreis2018, reinhardt2020, woo2022} and underscores the necessity of using realistic EOS models when assessing whether the giant impact scenario can reproduce the observed 30--50\% rock fraction of Uranus's satellites.
While these macroscopic differences are evident, it is important to distinguish whether they arise from the differences in the initial density profiles or from the varied thermodynamic responses during the impact.
We explore this dynamic and thermodynamic evolution in detail in Sections 4.3 and 4.4.
Furthermore, a comprehensive evaluation of how the specific shortcomings of each EOS (as described in Section 2.1) introduce specific biases into these impact dynamics is provided in Section 4.5.

\section{Discussion}

\subsection{Comparison with Previous Studies}

A detailed comparison of our results with earlier studies highlights the strong influence of the adopted equation of state (EOS) on the composition of the post-impact disk. 

For the SESAME/ANEOS EOS, our results are broadly consistent with those of \citet{slattery1992}, who reported that the impactor core was insufficiently disrupted, leaving little rocky or iron material in the debris. 
Similarly, our simulations show that while certain SESAME/ANEOS-S runs can produce disks containing more rock than the combined mass of Uranus's present satellites, the rock-to-ice ratio remains very low, typically below a few percent. 
Thus, SESAME/ANEOS tends to predict ice-dominated disks, even in cases with substantial total mass, and it is difficult to reconcile these outcomes with the observed 30--50\% rock fraction of Uranus's satellites \citep{jacobson1992masses}.
However, it should be noted that a direct comparison is somewhat limited since their models utilized a lower resolution of only $8 \times 10^3$ particles.

In contrast, our Tillotson EOS results differ from those of \citet{reinhardt2020}. 
Their simulations of a differentiated $3\,M_{\oplus}$ impactor produced virtually no rocky material in the disk, whereas our Tillotson-S runs—particularly at high angular momentum—yield disks containing sufficient rock mass relative to the present Uranian satellites, with rock fractions reaching $\sim$30--40\%. 
This discrepancy likely reflects differences in initial conditions: \citet{reinhardt2020} adopted a core-to-mantle ratio of 12:88, resulting in a much smaller initial reservoir of rocky material, whereas our setup incorporated a higher rock fraction that facilitated rock ejection. 
It may also be related to numerical differences in how each study treated impact geometry and the classification of disk particles, which can affect whether marginally bound rocky material is retained or reaccreted. 
Given that they also employed a DISPH-like formalism, the differences in outcomes may be more heavily influenced by the initial compositions rather than the numerical schemes.

Our results with the Tillotson EOS are instead more comparable to those of \citet{woo2022}, who showed that a disk with a rock fraction of $\sim$38\% could reproduce the composition of the present Uranian satellites. 
However, their study required extremely rock-rich impactors (up to 100\% rock) to achieve this outcome. 
By contrast, our results demonstrate that a more moderate rock fraction in the initial bodies, when coupled with the Tillotson EOS, can already yield disks with rock abundances in the observed range, at least under high-angular-momentum conditions. 
This contrast emphasizes that the predicted satellite composition is not only sensitive to the EOS, but also to assumptions about the impactor's initial structure determined by the chosen EOS and the definition of bound disk material. 

For the HM80 EOS, \citet{kegerreis2018} defined the ``disk'' as particles that are gravitationally bound and located outside the Roche radius of the proto-Uranus.
By comparing the right panel of our Appendix Figure~\ref{fig:DC_S_a}—which adopts the same definition—with Figure~9 of \citet{kegerreis2018}, we find that our simulations, unlike their $3M_{\oplus}$ case, produce a debris disk that is almost entirely devoid of rocky material.

This discrepancy primarily arises from the difference in the atmospheric mass retained by the target between the two studies.
Qualitatively, the mass of material ejected from the target is determined by the amount released directly during the impact, whereas the mass ejected from the impactor depends on the material expelled from the arm-like structure that forms after the collision.
Consequently, both Figure 9 of \citet{kegerreis2018} and our Figure \ref{fig:DC_S_a} exhibit the same overall trend: As the angular momentum increases, the amount of target-derived ice ($\mathrm{Ice_{tar}}$) decreases, while the amount of impactor-derived ice ($\mathrm{Ice_{imp}}$) increases.
While shock wave convergence can theoretically eject material at the planet's antipode, particle tracking analysis indicates that this mechanism primarily affects the extended H-He envelope.
For the rock and ice particles relevant to satellite formation, the antipodal contribution—conservatively defined as the entire far-side hemisphere—shows a dependence on the chosen EOS.
In the Tillotson and SESAME/ANEOS models, this contribution is minor, typically ranging from $0\% \to ~6.5\%$.
Although the HM80 EOS yields a higher antipodal contribution—typically between $5\%$ and $15\%$, and reaching up to $\sim25\%$ specifically in SSPH simulations with high impact angular momentum—the majority  of the condensable disk material originates from the disrupted impactor and the target's primary impact interface across all cases. 

However, because \citet{kegerreis2018} assumed a relatively thin atmosphere of $0.84 M_{\oplus}$, whereas our simulations adopted a much more massive envelope of $2 M_{\oplus}$, the impactor-derived particles in our models lose angular momentum more efficiently through interactions with the dense atmosphere and the forming arm structure.
As a result, two key differences emerge: (1) The amount of $\mathrm{Ice_{imp}}$ is reduced, preventing the reversal between $\mathrm{Ice_{imp}}$ and $\mathrm{Ice_{tar}}$ seen in \citet{kegerreis2018}$;$ and (2) no impactor-derived rock ($\mathrm{Rock_{imp}}$) is ejected in our simulations.

Taken together, these comparisons highlight that the composition of the debris disk—unlike its total mass or radial extent—is highly model-dependent. 
SESAME/ANEOS consistently predicts ice-dominated disks, while Tillotson EOS can produce rock-rich outcomes under certain conditions. 
Reconciling these divergent predictions remains a key challenge for evaluating the plausibility of the giant impact scenario for Uranus.

\subsection{Comparison between SSPH and DISPH}

\begin{figure}[ht!]
\includegraphics[width=\linewidth]{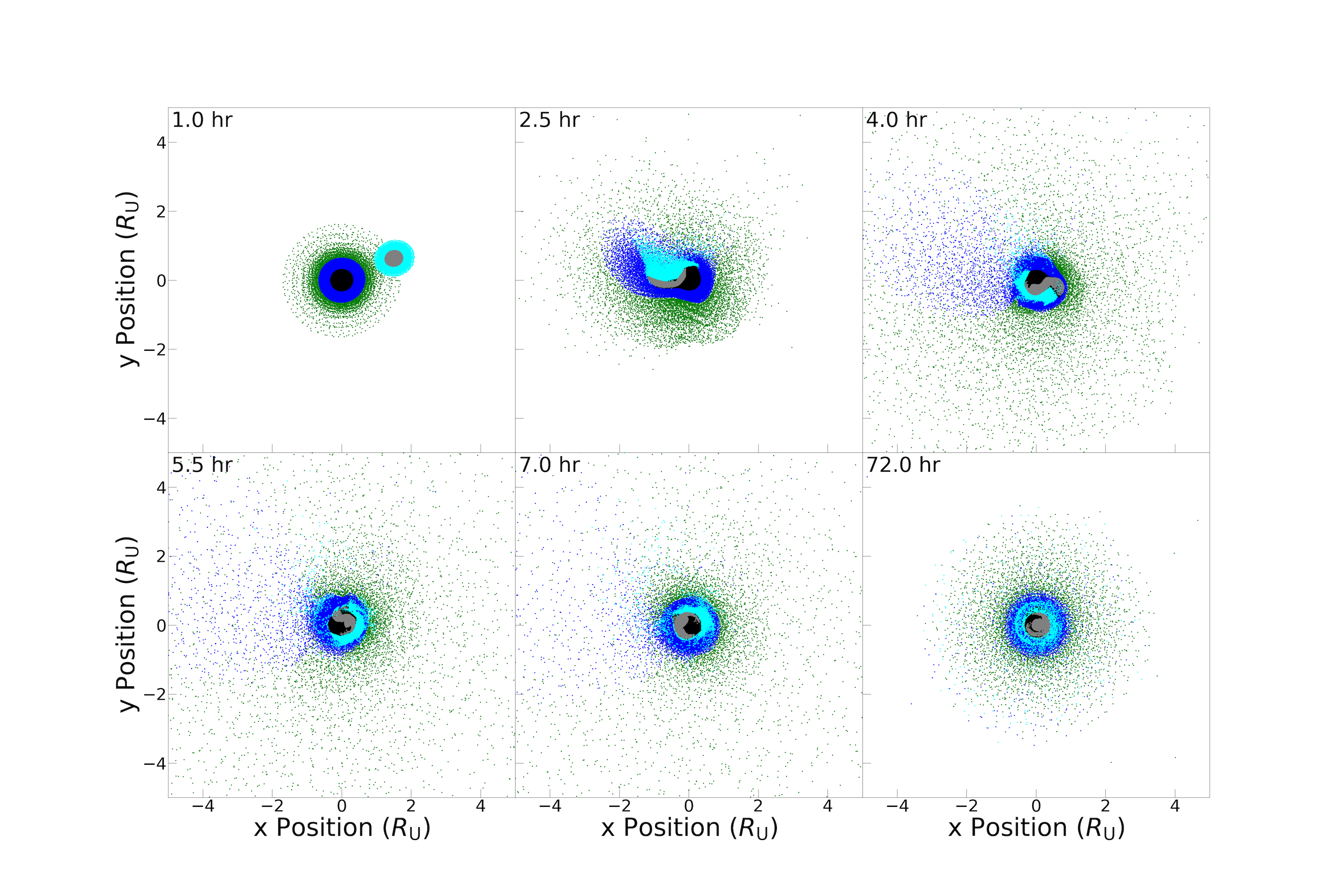}
\caption{Snapshots of a collision with a $L_{imp} = 3 \times 10^{36}\, {\rm kg \,m^2 \,s^{-1}}$ using SESAME and ANEOS EOS with SSPH. Particles with $z < 0$ are shown. The snapshot times are given in hours from the beginning of the simulation.}\label{figure:snapshot_SESAME_S_3}
\end{figure}

\begin{figure}[ht!]
\includegraphics[width=\linewidth]{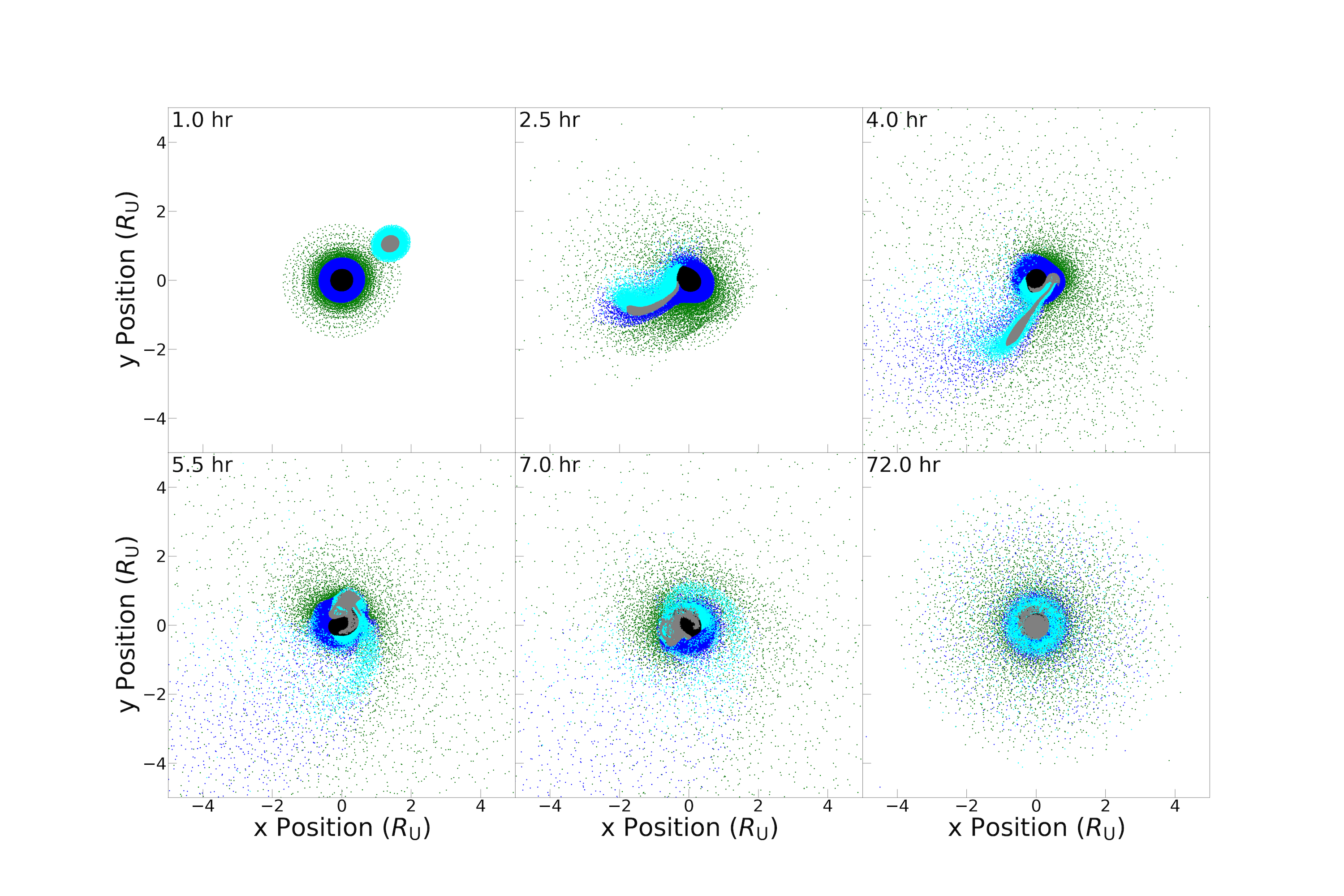}
\caption{The same as Fig. \ref{figure:snapshot_SESAME_S_3}, except for a low-impact parameter $L_{imp} = 5 \times 10^{36} {\rm kg \,m^2 \,s^{-1}}$.}\label{figure:snapshot_SESAME_S_5}
\end{figure}

\begin{figure}[ht!]
\includegraphics[width=\linewidth]{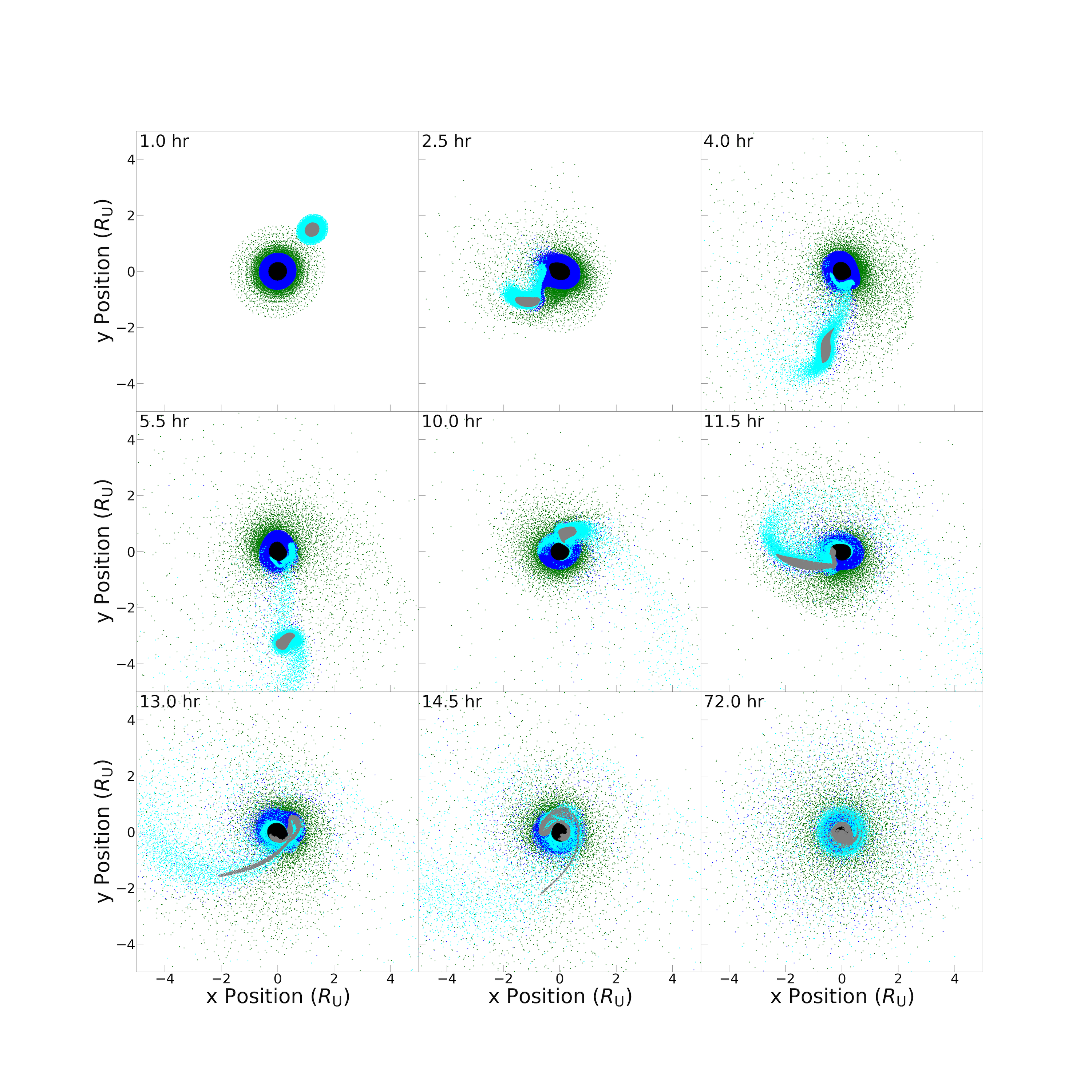}
\caption{The same as Fig. \ref{figure:snapshot_SESAME_S_3}, except for a high-impact parameter $L_{imp} = 7 \times 10^{36} {\rm kg \,m^2 \,s^{-1}}$.}\label{figure:snapshot_SESAME_S_7}
\end{figure}

Figures~\ref{figure:snapshot_SESAME_S_3}–\ref{figure:snapshot_SESAME_S_7} show the temporal evolution of collisions with $L_{\rm imp} = 3.0$, $5.0$, and $7.0 \times 10^{36}\,{\rm kg\,m^2\,s^{-1}}$ using the SESAME/ANEOS EOS with the SSPH scheme.
Figures~\ref{fig:DiskMass} and~\ref{fig:DiskSize} demonstrate that the differences between the SSPH and DISPH schemes do not significantly affect the overall mass or size of the resulting debris disk.

Although \citet{hosono2016} reported that switching from SSPH to DISPH generally produces a more compact disk, such a trend is not observed in our calculations.
We hypothesize that this discrepancy arises because the differences between the two SPH formulations are masked within the low-density atmosphere, where the numerical accuracy of SPH is inherently limited.
Since the behavior of this tenuous atmosphere predominantly governs the outer extent of the disk, the numerical distinction between SSPH and DISPH becomes less apparent in our models.

In contrast, Figures~\ref{fig:DC_S} and~\ref{fig:DC_DI} indicate that transitioning from SSPH to DISPH clearly suppresses the ejection of rocky material into the disk.
This difference likely results from the fact that DISPH alleviates the issue of non-physical surface tension that plagues SSPH.
Such artificial tension originates from the improper treatment of contact discontinuities—particularly at the core–mantle boundary and planetary surface—and is a well-known limitation of the standard SPH method.

For the HM80 EOS, the disk composition shows that, when using SSPH, the amount of target-derived ice exceeds that of impactor-derived ice (Figures~\ref{fig:DC_S} and~\ref{fig:DC_DI}).
However, this balance reverses when the DISPH scheme is employed.
\citet{hosono2016} demonstrated that in SSPH, the inaccurate handling of contact discontinuities generates a spurious repulsive pressure at the core–mantle boundary during impact, which drives material outward more strongly.
A similar mechanism appears in our SSPH simulations, causing the dominance of target-derived ice over impactor-derived ice.
The reversal seen in DISPH therefore reflects the ability of the DISPH formulation to eliminate this artificial repulsive force and to model core–mantle interactions more realistically.

\subsection{The Role of Initial Internal Structure}
\begin{figure}[ht!]
\includegraphics[width=\linewidth]{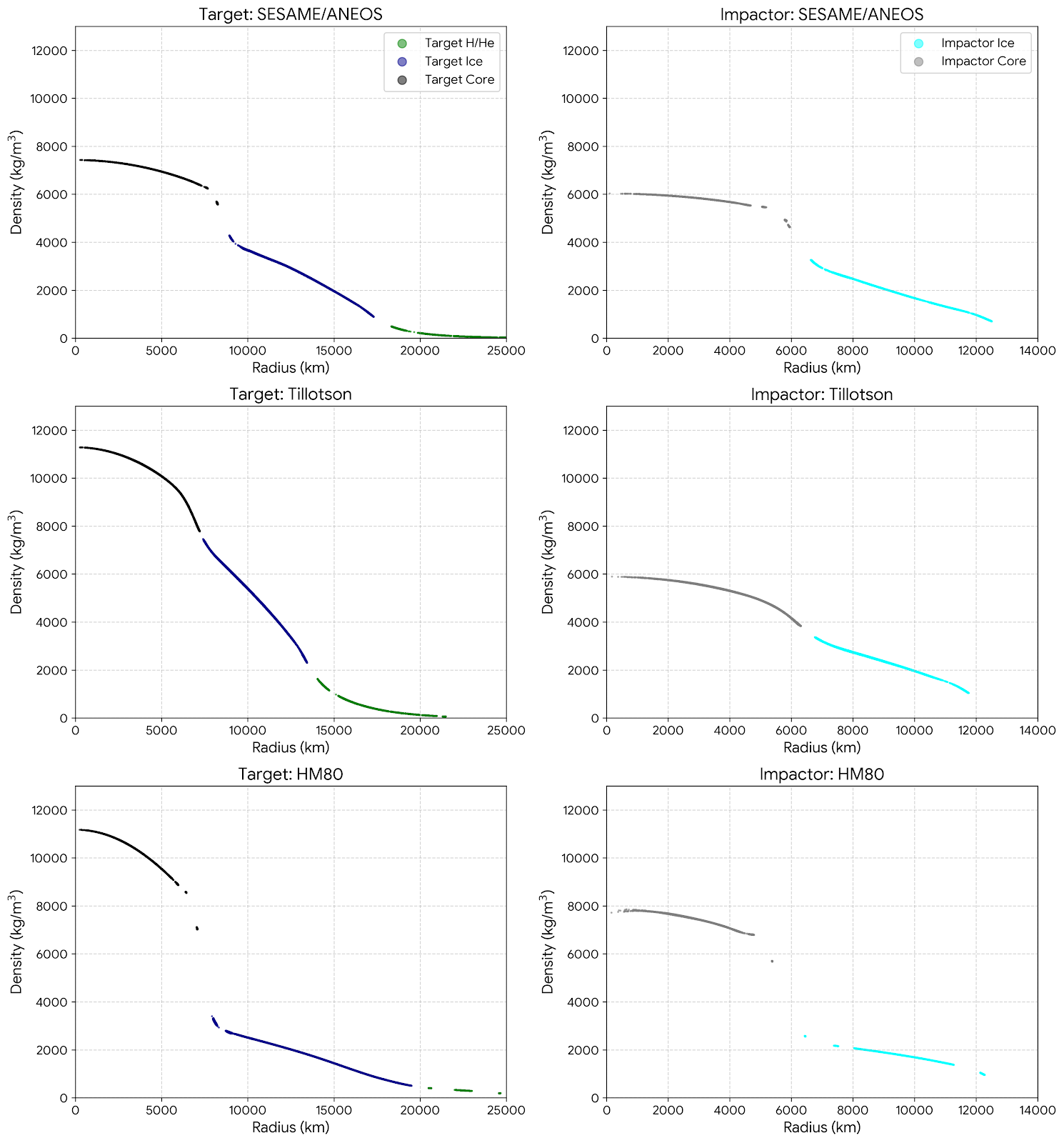}
\caption{Initial density profiles of the target (left column) and the impactor (right column) utilizing the SSPH. From top to bottom, the panels show the profiles generated by the SESAME/ANEOS, Tillotson, and HM80 equations of state. The radial distance is calculated from the center of mass of each respective body at $t = 0$. }\label{figure:IC_S}
\end{figure}

\begin{figure}[ht!]
\includegraphics[width=\linewidth]{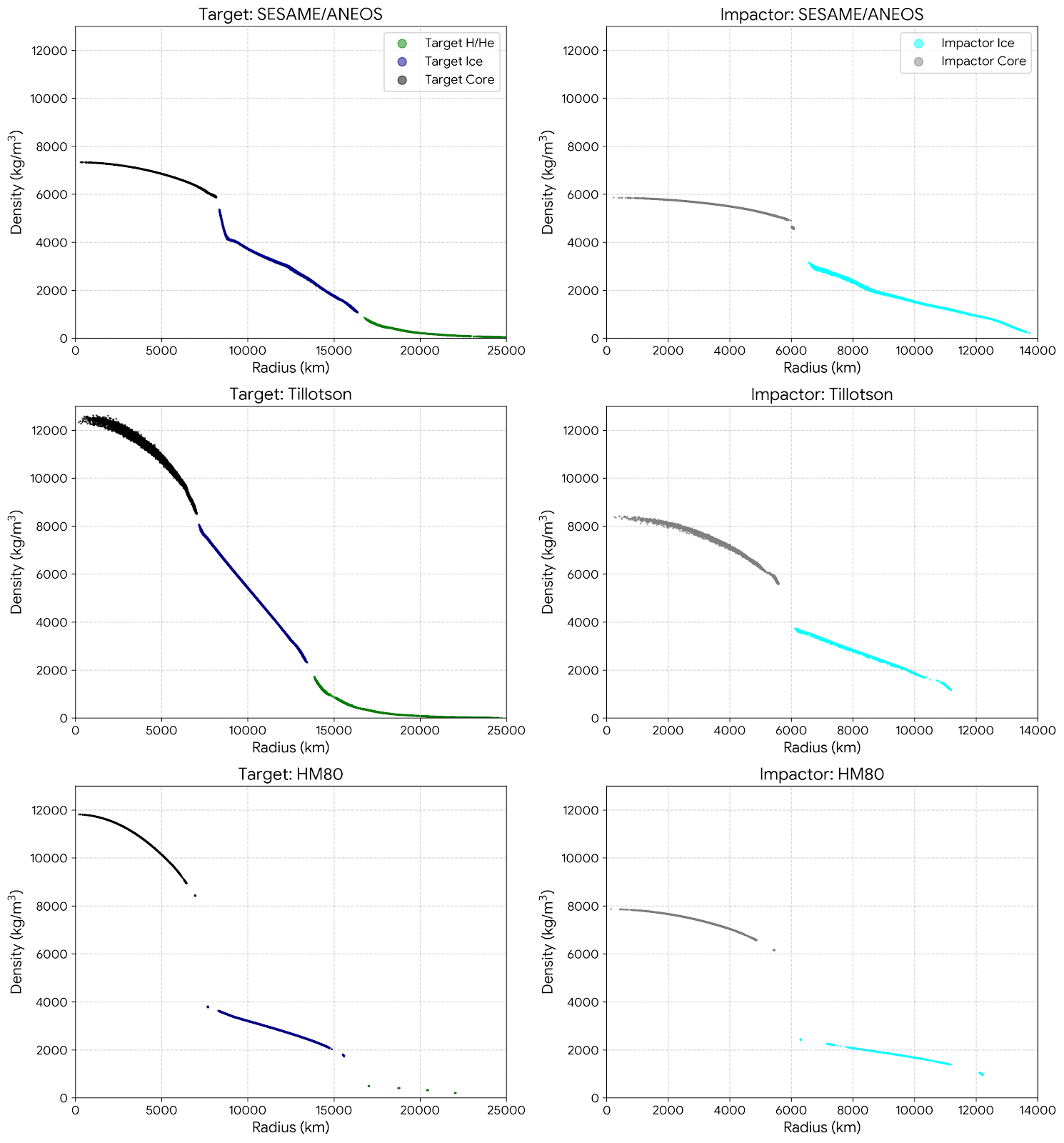}
\caption{Initial density profiles of the target (left column) and the impactor (right column) utilizing the DISPH. The layout and the EOS arrangements are identical to those in Fig. \ref{figure:IC_S}}\label{figure:IC_DI}
\end{figure}

To understand the origin of the compositional differences in the debris disks across different EOS models, it is essential to consider the initial internal structure of the colliding bodies.
Although the global mass and the bulk composition are strictly identical across all impact scenarios in this study, the different compressibility of materials handled by each EOS inherently alters the initial density distributions.
This intrinsic variation is clearly evident in the initial density profiles for both the density-independent SPH (DISPH) and standard SPH (SSPH) setups (see Figures~\ref{figure:IC_S} and ~\ref{figure:IC_DI}).

A physical distinction among the models lies in the magnitude of the density discontinuity at the core-mantle boundary.
In configurations utilizing the HM80 and SESAME/ANEOS EOS, a pronounced density jump inherently forms between the compact silicate core and the surrounding ice mantle.
In stark contrast, the Tillotson EOS dictates a much smoother density transition across this boundary.

This inherent structural difference fundamentally dictates the primary excavation phase during a grazing impact.
A sharp density discontinuity, such as that produced by the HM80 and SESAME/ANEOS models, acts as a strong dynamic barrier—an impedance mismatch.
During the initial collision, while the outer ice mantle is heavily deformed and stripped away by shear forces, the distinct, high-density rocky core naturally resists being dragged outward.
Conversely, the absence of a sharp density jump in the Tillotson EOS models leaves the core and mantle more continuously coupled.
Lacking this mechanical barrier, the underlying rocky core material is easily excavated and dynamically dragged outward into the initial spiral arms along with the expanding ice.

\subsection{Secondary Impacts}

The differences among EOSs become most pronounced in cases where a secondary collision occurs. 
Such events arise when material ejected during the first encounter remains marginally bound and subsequently re-impacts the planet on timescales of several hours. 
Similar multiple-impact sequences have been reported in other SPH studies of Uranus-scale collisions \citep[e.g.,][]{kegerreis2018, reinhardt2020}, and they may play an important role in redistributing material between the disk and the planet. 
We therefore examined the details of such second impacts in our simulations. 

\begin{figure}[t]
\includegraphics[width=\linewidth]{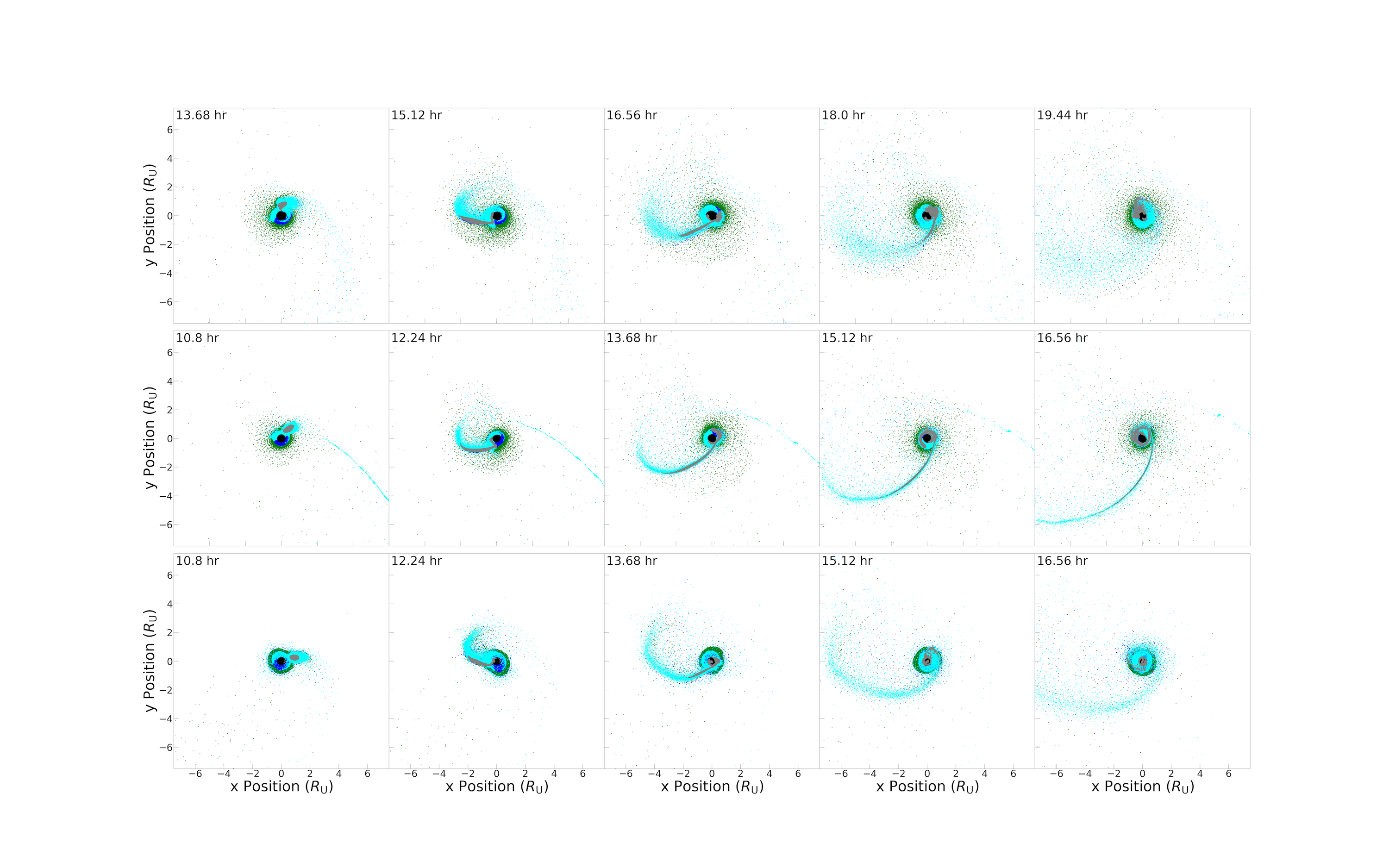}
\caption{Snapshots of the second collision with DISPH and $L_{\rm imp} = 7.0 \times 10^{36} {\rm kg \,m^2\,s^{-1}}$ for SESAME/ANEOS (top), Tillotson EOS (middle), and HM80 EOS (bottom).}\label{fig:SI}
\end{figure}

\begin{figure}[t]
\includegraphics[width=\linewidth]{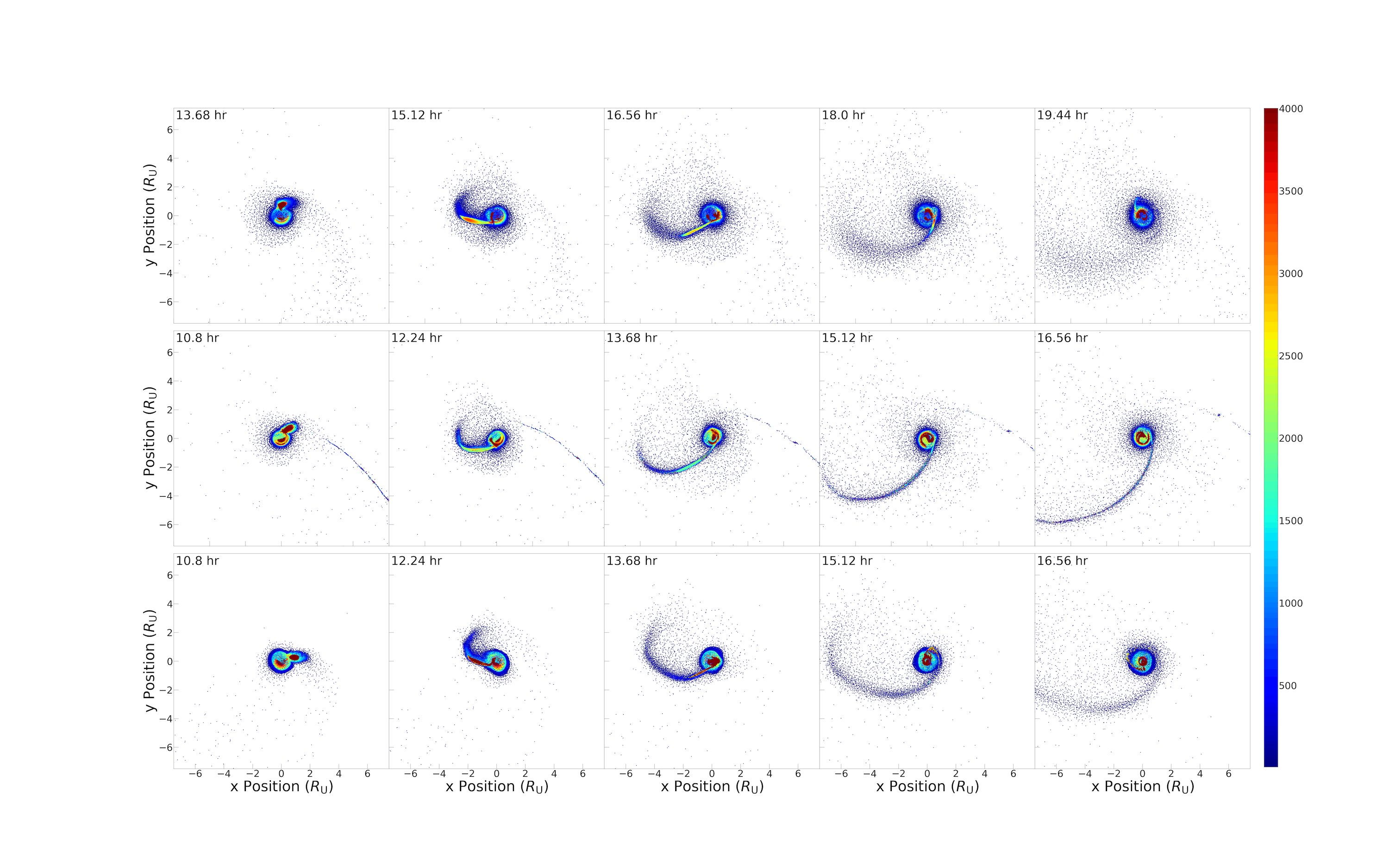}
\caption{The same results as Figure~\ref{fig:SI}, but with density represented by the color scale.}\label{fig:SI_dens}
\end{figure}

\begin{figure}[t]
\includegraphics[width=\linewidth]{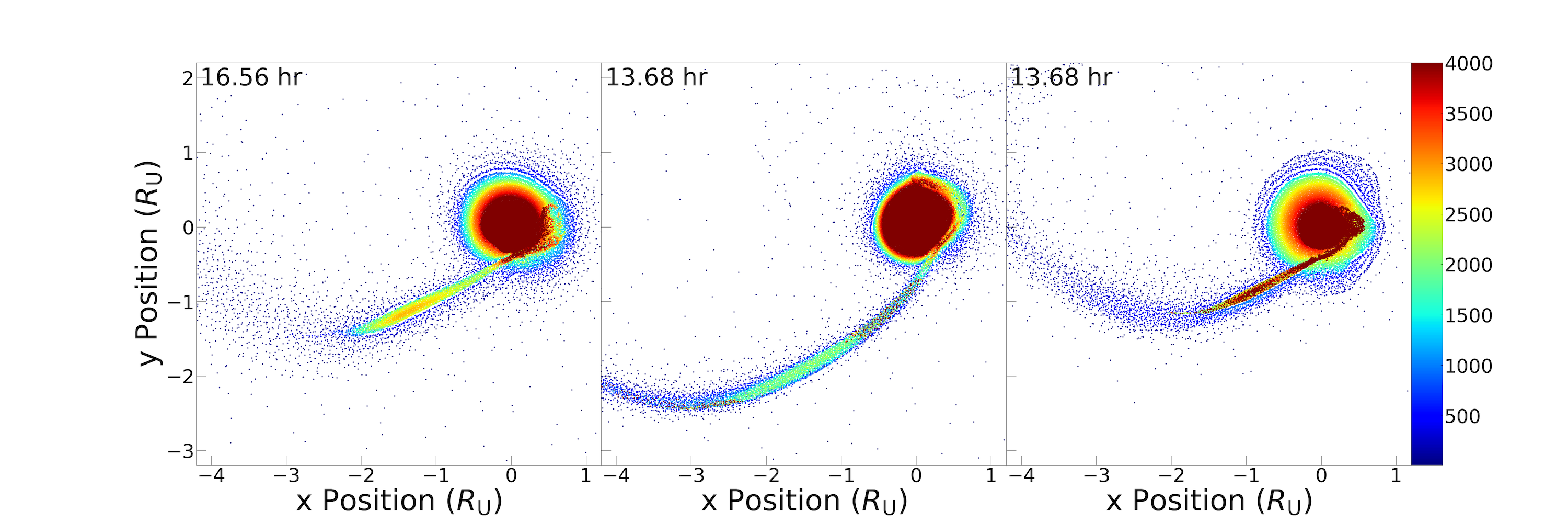}
\caption{Density map of snapshots of second collision for SESAME/ANEOS (left), Tillotson EOS (center), and HM80 EOS (right).}\label{fig:dens_detail}
\end{figure}

Figure~\ref{fig:SI} shows snapshots of the second collision for each EOS, and Figure~\ref{fig:SI_dens} is the density map of Figure~\ref{fig:SI}.
Figure~\ref{fig:dens_detail} presents the corresponding density maps for simulations performed with DISPH at $L_{\rm imp} = 7.0 \times 10^{36}\,{\rm kg\,m^2\,s^{-1}}$.
The overall outcomes of the SESAME/ANEOS and Tillotson models are qualitatively similar, whereas those of the HM80 EOS differ substantially. 
Because the atmospheric EOS is treated differently only in HM80, the atmosphere is not efficiently dispersed by the impact. 
As a result, the arm-like structures formed from the impactor remain suspended in the envelope rather than falling back onto the planet, owing to the reduced atmospheric drag. 
This produces a more extended, quasi-stable debris configuration compared to the other two EOSs. 

Comparing SESAME/ANEOS with Tillotson, the density of rock in the arms formed during the second collision is systematically higher in the SESAME/ANEOS case (Fig.~\ref{fig:dens_detail}). 
Consequently, for SESAME/ANEOS (and also for HM80) the rocky clumps rapidly coalesce and fall back onto Uranus, depleting the disk of heavy material within a few hours. 
By contrast, in the Tillotson case the arms spread out more diffusively, and a significant fraction of the rocky component remains dispersed across the disk for extended timescales. 
Quantitatively, the amount of rock retained in the disk can exceed $\sim 10^{2}\,M_{\rm satellite}$ in high-$L_{\rm imp}$ Tillotson runs, whereas SESAME/ANEOS cases often retain less than $\sim 10\,M_{\rm satellite}$ after the second impact. 

This contrast highlights how EOS-dependent treatments of both the atmosphere and the rocky interior strongly influence the redistribution of material in successive impacts. 
In particular, the absence of latent heat in Tillotson facilitates rock ejection and survival in the disk, while the more realistic phase transitions included in SESAME/ANEOS promote re-accretion of rocky material. 
The HM80 EOS, by maintaining a dense envelope that suppresses fallback, yields yet another distinct outcome. 

These findings suggest that secondary impacts are a potentially critical factor for understanding the eventual rock-to-ice ratio of the circum-Uranian disk. 
Depending on the EOS, the same initial giant impact can either strip the disk of rocky clumps (SESAME/ANEOS), retain them in orbit (Tillotson), or suspend them in the envelope (HM80). 
Thus, the treatment of material thermodynamics and atmospheric structure may control not only the total mass of the disk, but also the fate of its rocky component during multi-stage impact events.

Therefore, we conclude that the stark differences in the final disk composition are the result of a synergistic two-stage process.
The EOS-driven initial structural difference determines the primary efficiency of rock excavation, while the thermodynamic capability of the EOS to model the expanded states dictates whether the excavated rock reaccretes or is successfully retained in the disk.

It should be noted that the number and mass distribution of these re-impacting fragments can be sensitive to the numerical resolution.
The precise physical reality of such secondary clumps, and the extent to which their formation and subsequent impacts are influenced by numerical effects, warrants further investigation in future higher-resolution studies.

\subsection{Potential Influences of EOS Limitations on Impact Dynamics}
Given the structural and thermodynamic differences observed among the models, the inherent limitations of each EOS described in Section 2 may have influenced the resulting disk compositions.

For the Tillotson EOS, the inability to treat latent heat might contribute to an over-expansion of the vaporizing ice mantle upon shock release \citep[e.g.][]{wissing_hobbs_2020, meier_etal_2021}.
Furthermore, its simplified analytical formulation may hinder the formation of a sharp density jump at the core-mantle boundary.
It is possible that these factors combined could lead to an overestimation of rock excavation and its subsequent retention in the circumplanetary disk.

Conversely, the extreme density discontinuity observed in the HM80 initial models might be partially influenced by the uncertainty of extrapolating its analytical formula into impact-generated thermodynamic regimes, as cautioned in Section 2.1.3.
If this density jump is artificially amplified by the extrapolation, it could act as an overly strong dynamic barrier, potentially leading to an underestimation of the excavated rock mass.

Similarly, while the SESAME and ANEOS models handle multiphase thermodynamics more comprehensively, the interpolation within their tabular grids has the potential to over-stabilize the liquid-vapor mixed phases \citep[e.g.][]{stewart_shock_2019}.
This numerical stabilization could potentially enhance the cohesive nature of the clumps artificially during secondary impacts, which might lead to an overestimation of the rock reaccretion efficiency.

Therefore, it is reasonable to suggest that the actual rock-to-ice ratio of the proto-Uranian disk might fall somewhere between the highly rock-depleted outcomes of the HM80/SESAME models and the rock-enriched outcomes of the Tillotson model.
Future simulations incorporating more advanced multiphase EOS models would be beneficial to further constrain these dynamics.

\subsection{Constraints on Formation Scenarios}

The major Uranian satellites are estimated to have rock-to-ice ratios close to 1:1 \citep{jacobson1992masses}.
In various interior models, Uranus is represented as ice-rich, with ices constituting most of its heavy element mass \citep[e.g.,][]{podolak1991interior, hubbard1995interior}.
Although some studies suggest alternative structures, such as higher rock fractions or compositional gradients \citep[e.g.,][]{helled2020uranus, vazan2020uranus}, ice-rich compositions are frequently used as the baseline for giant impact simulations.
Assuming that proto-Uranus and the involved impactors have similar ice-rich bulk compositions, it remains to be determined whether the giant impact scenario can produce a satellite system that is relatively enriched in rock compared to Uranus.
It should be noted, however, that the formation and migration history of the ice giants remains poorly constrained \citep[e.g.,][]{tsiganis2005origin, desch2018formation}, implying that the impactor could have originated from a different region of the solar nebula with a different initial rock-to-ice ratio.
Utilizing identical bulk compositions serves as a controlled baseline to isolate the dynamical effects of the collision.

As shown in Figure~7, for collisions with high initial angular momentum the total mass of rock incorporated into the debris disk is tens of times larger than the combined mass of Uranus's current satellites, largely independent of the adopted EOS. 
However, in the case of the SESAME/ANEOS model—the most physically realistic EOS considered in this study—the rock fraction of the disk remains well below the inferred 30--50\% value, typically $\lesssim 5\%$. 
This finding is consistent with earlier results \citep[e.g.,][]{slattery1992, kegerreis2018, reinhardt2020}, which also reported ice-dominated disks when realistic EOS models were applied. 
By contrast, the Tillotson EOS yields rock fractions of $\sim $30--40\% under favorable conditions, but this result reflects its neglect of latent heat and simplified treatment of phase transitions rather than a physically realistic outcome. 

The discrepancy between SESAME/ANEOS predictions and the observed satellite compositions suggests that additional processes must have operated after the initial impact. 
Several possible mechanisms have been proposed: (i) Impacts involving intrinsically rock-rich impactors, which can enhance the rocky component of the disk \citep{woo2022}; (ii) subsequent contamination or accretion of rocky debris from the Kuiper Belt or planetesimal population; and (iii) preferential enrichment of rock during the thermodynamic evolution of the vapor disk. 
In particular, \citet{ida2020} implied that the rock/ice ratio of condensed solids can become substantially higher than that of the vapor–liquid mixture in the impact-generated disk, owing to the higher condensation temperatures of silicates compared to ices. 
Such enrichment could naturally elevate the rock fraction of the building blocks of satellites even if the initial disk was ice-dominated. 
Furthermore, as an alternative explanation to the impact-generated vapor disk model, \citet{reynard_sotin_2024} analyzed the densities of Uranus's moons and proposed that their current ice-to-rock ratios could be evidence of ice/rock fractionation processes that occurred during the planetary accretion phase.
This provides a distinct mechanism for rock enrichment without strictly relying on the thermodynamic evolution of an impact debris disk.

Furthermore, a fundamental challenge for the giant impact hypothesis that must be acknowledged is the excess angular momentum (AM) problem.
Single giant impacts that are energetic and oblique enough to reproduce Uranus's ~98° axial tilt typically impart a total AM to the post-impact system (planet plus disk) that exceeds the present-day angular momentum of the Uranian system \citep[e.g.,][]{slattery1992, morbidelli2012}.
Resolving this discrepancy remains a major open issue in the field. It implies that the system must have experienced significant AM loss during its subsequent evolution—potentially through mass shedding from the extended, rotating envelope or strong disk-planet interactions—or that Uranus's tilt was instead produced by a different dynamical pathway, such as a sequence of multiple smaller impacts \citep{morbidelli2012}.
While the present study focuses primarily on the dynamical characteristics (e.g., mass and size) and the bulk composition of the impact-generated disk, these broader constraints highlight the necessity of understanding the long-term angular momentum evolution of the post-impact system.

Our findings therefore reinforce the importance of post-impact disk evolution in determining the final composition of the Uranian satellites. 
While \citet{woo2022} pioneered the end-to-end coupling of hydrodynamic impact outcomes with 1D disk evolution and N-body accretion models, as noted in their study, their framework relied on simplified condensation assumptions (e.g., constant condensation temperatures) and required an extreme initial condition—a 100\% pure rocky impactor—to reproduce the observed 1:1 rock-to-ice ratio.
In future work, we plan to significantly improve upon these foundational efforts by coupling impact simulations with more advanced, pressure-dependent disk thermodynamic models that include detailed condensation physics and phase separation.
Such refined coupled models will be essential to rigorously assess whether the satellite system's composition can be naturally reproduced from more physically plausible, ice-rich precursors.
Only by incorporating both the impact dynamics and the subsequent evolutionary processes can the plausibility of the giant impact scenario for Uranus be fully constrained.

\section{Summary}

We performed SPH simulations of giant impacts onto proto-Uranus using two numerical schemes (SSPH and DISPH) and three different equations of state (ANEOS/SESAME, Tillotson EOS, and HM80 EOS). 
Our goal was to investigate how the outcome of such impacts—including the rotation period of the planet, the debris disk mass $M_{\rm d,imp}$, its characteristic radius $\langle r_{\rm d,imp}\rangle$, and the rock fraction in the disk—depends on the choice of EOS and SPH scheme. 

Previous SPH studies have often predicted debris disks that were too massive and compact compared to Uranus's present satellite system. 
\citet{ida2020} demonstrated that viscous spreading of a fully vaporized disk could reconcile the disk mass and size with the current system, but the ice-to-rock ratio remained uncertain. 
Our analysis provides a detailed assessment of these issues and leads to the following conclusions:

\begin{itemize}
    \item For collisions capable of reproducing the current Uranian system, the influence of EOS and SPH scheme on the final planetary rotation period, as well as on the overall disk mass and size, is relatively small.
    Furthermore, resolution tests conducted up to $10^6$ particles using the Tillotson EOS (Appendix B) suggest that these macroscopic outcomes are relatively insensitive to numerical resolution within this range. However, it should be noted that further increasing the resolution could potentially alter the results.
    \item The choice of SPH scheme (SSPH vs. DISPH) has only a minor effect on the rock mass fraction in the disks, indicating that composition is determined primarily by the EOS rather than by the numerical formulation. 
    However, the choice of SPH scheme does introduce subtle differences in the compositional details of the disk.
    Notably, the DISPH scheme generally suppresses rock ejection compared to the SSPH scheme.
    This difference arises because the DISPH formulation eliminates the artificial surface tension at the core-mantle boundary, thereby preventing the spurious repulsive forces that unphysically exaggerate rock ejection in SSPH.
    Consequently, the efficiency of rock deposition is primarily governed by the chosen EOS rather than the SPH scheme, with the Tillotson EOS yielding substantial rock fractions while the SESAME and ANEOS models produce ice-dominated disks.
    \item For the Tillotson EOS, the disk mass can reach $\sim$ 100 times the total mass of the present Uranian satellites, with rock fractions of 30--40\%, broadly consistent with the observed satellite system. 
    By contrast, with SESAME/ANEOS and HM80 EOSs, the disk masses are typically only $\sim$ 1--10 times larger than the present satellites, and the rock fraction remains $\lesssim 5\%$. 
    The relatively high rock fraction obtained with the Tillotson EOS likely reflects both the assumed internal structure of the impactor/target and the neglect of latent heat and phase transitions in this EOS. 
\end{itemize}

In particular, the Tillotson-S model yields disks with rock fractions exceeding 40\%. 
\citet{woo2022} showed that a disk with $\sim$ 38\% rock can reproduce the present satellite system, although their study employed highly rock-rich impactors and incorporated additional processes not considered here. 
Our results demonstrate that such rock-rich outcomes can be achieved under less extreme assumptions, but only when simplified EOSs are used. 

By contrast, our results with SESAME/ANEOS—the most realistic EOS employed, as it explicitly accounts for phase transitions—indicate that it is difficult to reproduce the present Uranian satellite system within a simple giant impact framework, consistent with earlier work \citep[e.g.,][]{slattery1992, kegerreis2018}. 
To explain the observed rock-to-ice ratios of Uranus's satellites, additional processes are therefore required, such as impacts involving intrinsically rock-rich impactors, subsequent contamination from Kuiper Belt objects, or preferential rock enrichment during the thermodynamic evolution of the vapor disk \citep{ida2020}. 

Overall, our study clarifies the origin of previously reported discrepancies in Uranus impact simulations by explicitly disentangling the effects of EOS and SPH scheme. 
It underscores that while disk mass and size are relatively robust to modeling choices, the rock fraction—and therefore the ability to reproduce the present Uranian satellites—is highly sensitive to the adopted EOS. 

Finally, we note the numerical limitations of the current study.
A baseline resolution of $10^5$ particles and a broad search radius (Wendland C6 kernel) were employed to maintain numerical stability and prevent pairing instability across extreme density contrasts. 
While validation tests (including $10^6$-particle and C2 kernel runs detailed in Appendix B) indicate that macroscopic quantities, such as the total disk mass and bulk composition, remain consistent, this setup limits the effective spatial resolution.
This configuration can result in the over-smoothing of sharp phase boundaries, which restricts the ability to trace the fine-scale mixing of rocky components.
Future high-resolution simulations are necessary to further quantify their detailed spatial distribution.
Furthermore, in our current SPH implementation, the smoothing length of each particle is defined to be proportional to its physical radius.
While the smoothing length varies in time and space according to the spatial extent of individual particles, it does not adjust to maintain a constant number of neighbors.
Consequently, in highly compressed regions, an excessive number of interacting neighbors can lead to over-smoothing of pressure gradients, while in expanding vapor clouds, a deficiency of neighbors may result in increased numerical noise or decoupling.
Implementing a fully adaptive smoothing length algorithm that explicitly regulates the neighbor count is a subject for future work to improve the resolution of both shock dynamics and low-density structures.

\begin{acknowledgments}
This work was supported by JSPS KAKENHI Grant Number 21H04512, 25K07384, and Grant-in-Aid for JSPS Research Fellow Number 22J22428.
We would like to express our gratitude to the reviewer for their thoughtful comments, which helped us refine our discussion on the giant impact hypothesis and the EOS models.
\end{acknowledgments}

\begin{appendix}

\section{Comparison with an Alternative Classification of SPH Particles}
\label{appendix_A}

\begin{figure}[t]
\includegraphics[width=\linewidth]{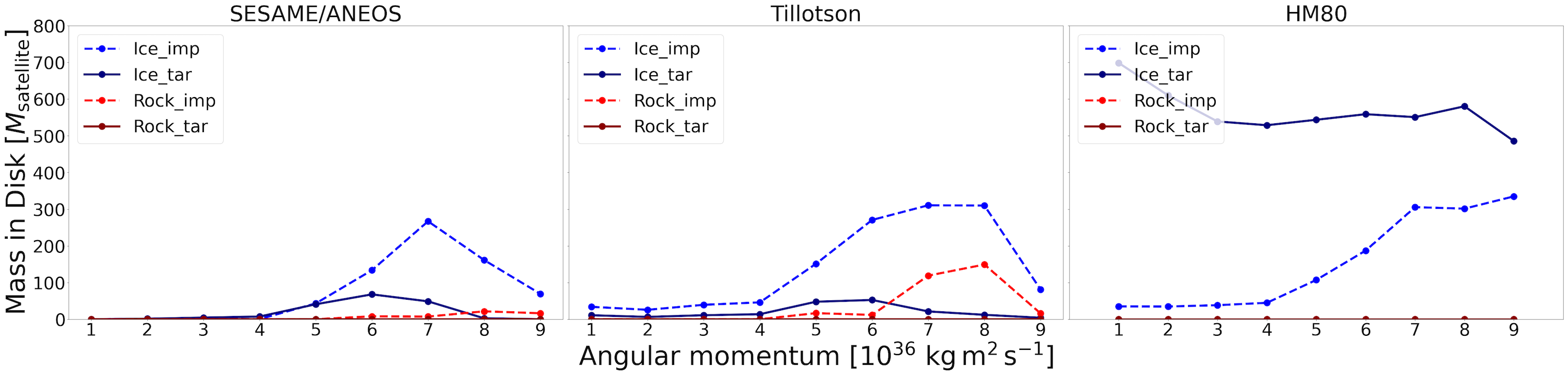}
\caption{Rock and ice components in the disk as a function of $L_{\rm imp}$ for different EOSs in SSPH (Appendix version). The total mass of rock components is normalized by the present satellite mass $M_{\rm satellite}$.}\label{fig:DC_S_a}
\end{figure}

\begin{figure}[t]
\includegraphics[width=\linewidth]{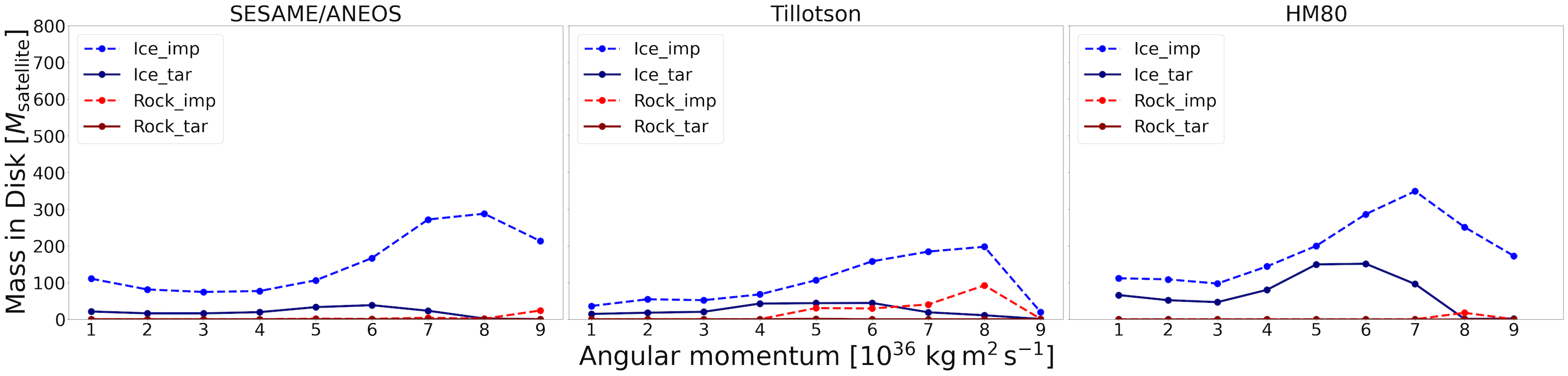}
\caption{Rock and ice components in the disk as a function of $L_{\rm imp}$ for different EOSs in DISPH (Appendix version). The total mass of rock components is normalized by the present satellite mass $M_{\rm satellite}$.}\label{fig:DC_DI_a}
\end{figure}

To investigate the properties of the resulting debris disks, SPH particles in this study were classified into three categories: ``planetary,'' ``disk,'' or ``unbound.''
The main text (Section~\ref{sec:results}) adopts the classification scheme of \citet{canup2001}, whereas the Appendix presents results based on a simpler definition in which disk particles are defined as gravitationally bound and located outside the Roche radius.
Figures~\ref{fig:DC_S_a} and~\ref{fig:DC_DI_a} provide alternative representations of the data shown in Figures~\ref{fig:DC_S} and~\ref{fig:DC_DI}, employing this simpler disk definition.
Note that the vertical-axis scales differ between these figures.

Although both definitions yield qualitatively similar trends, the iterative algorithm of \citet{canup2001} generally produces a smaller disk mass, as it systematically excludes particles that are only weakly bound or are likely to reaccrete onto the planet over short timescales.

\section{Numerical Convergence on Resolution and Kernel Choice}
\label{sec:appendixB}

\begin{figure}[t]
\includegraphics[width=\linewidth]{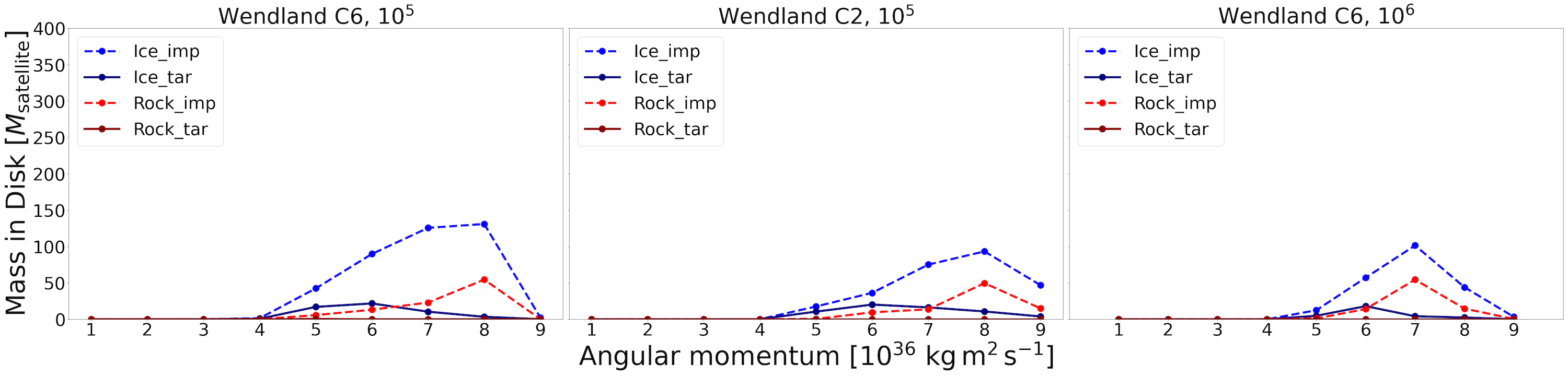}
\caption{Rock and ice components in the disk as a function of $L_{\rm imp}$. Left panel: Baseline simulations utilizing the Wendland C6 kernel and approximately $10^5$ particles. Middle panel: Test for the kernel choice using the Wendland C2 kernel with $N \approx 10^5$ particles. Right panel: Test for spatial resolution using $N \approx 10^6$ particles and the Wendland C6 kernel.}\label{fig:AppendixB}
\end{figure}

To evaluate the numerical robustness of our results, we performed supplementary simulations varying the SPH kernel and the total number of particles. 
Figure~\ref{fig:AppendixB} shows the mass of each component (ice, impactor-derived rock, and target-derived rock) within the debris disk as a function of the initial angular momentum $L_{\text{init}}$ under three different numerical configurations: our baseline setup (Wendland C6, $N \approx 10^5$), a sensitivity test for the smoothing kernel (Wendland C2, $N \approx 10^5$), and a convergence test for spatial resolution ($N \approx 10^6$).

While minor discrepancies are observed between these cases—such as slight shifts in the peak location of ejected rock mass or small variations in the total ice mass—the overall trends remain remarkably consistent.
Specifically, the qualitative dependence of the disk composition on the impact parameters and the partitioning of different materials are well-captured across all tested kernels and resolutions.
However, we note a limitation regarding the assessment of spatial convergence.
While increasing the particle number from $10^5$ to $10^6$ improves the mass resolution by an order of magnitude, the corresponding gain in spatial resolution scales as $h \propto N^{-1/3}$ (a factor of $\approx 2.15$).
This scaling, combined with the use of a constant smoothing length, makes it difficult to separate the effects of improved spatial resolution from those of increased kernel sampling.
Therefore, while these supplementary results indicate that our baseline setup reproduces the overall macroscopic trends, future simulations utilizing higher resolutions and adaptive kernels are required to further evaluate numerical convergence.

\end{appendix}

\bibliography{reference}

\end{document}